\documentclass[12pt]{article}

\usepackage[top=80pt,bottom=85pt,left=67pt,right=67pt]{geometry}
\usepackage{amssymb,amsmath}
\usepackage{graphicx,color}
\usepackage{float}
\usepackage{dsfont}
\usepackage{cite}
\usepackage[debug,pageanchor=false]{hyperref}
\definecolor{link}{rgb}{.8,.15,.1}
\hypersetup{colorlinks=true,linkcolor=link,citecolor=link,urlcolor=link,linktocpage}

\makeatletter
\@addtoreset{equation}{section}
\makeatother

\renewcommand{\theequation}{\thesection.\arabic{equation}}

\def\hat{\widehat}

\newcommand{\beq}{\begin{equation}}
\newcommand{\eeq}{\end{equation}}
\newcommand{\bea}{\begin{eqnarray}}
\newcommand{\eea}{\end{eqnarray}}
\newcommand{\nn}{\nonumber}

\begin{document}

\begin{titlepage}

\begin{center}

\vskip .9in 
\noindent

{\Large \bf{$M$-strings and AdS$_3$ solutions to M-theory with small $\mathcal{N}=(0,4)$ supersymmetry}}

\bigskip\bigskip

Yolanda Lozano$^{a,}$\footnote{ylozano@uniovi.es},  Carlos Nunez$^{b,}$\footnote{c.nunez@swansea.ac.uk}, Anayeli Ramirez$^{a,}$\footnote{anayelam@gmail.com}, Stefano Speziali$^{b,}$\footnote{stefano.speziali6@gmail.com} \\

\bigskip\medskip
{\small

 $a$: Department of Physics, University of Oviedo,
Avda. Federico Garcia Lorca s/n, 33007 Oviedo, Spain
\vskip 3mm
 $b$: Department of Physics, Swansea University, Swansea SA2 8PP, United Kingdom.}

\vskip 2cm 
     	{\bf Abstract }

\vskip .1in
\end{center}

\noindent
We construct a general class of (small) $\mathcal{N}=(0,4)$ superconformal solutions in M-theory  of the form AdS$_3\times S^3/\mathbb{Z}_k\times \text{CY}_2$,
foliated over an interval. These solutions describe M-strings in M5-brane intersections. The $M$-strings support $(0,4)$ quiver CFTs that are in correspondence with our backgrounds. We compute the central charge and show that it scales linearly with the total number of $M$-strings. We introduce momentum charge, thus allowing for a description in terms of M(atrix) theory. Upon reduction to Type IIA, we find a new class of solutions with four Poincar\'e supercharges of the form AdS$_2\times S^3\times \text{CY}_2\times \mathcal{I}$, that we extend to the massive IIA case. We generalise our constructions to provide a complete class of AdS$_3$ solutions to M-theory with (0,4) supersymmetry and SU(2) structure. We also construct new AdS$_2\times S^3\times \text{M}_4\times \mathcal{I}$ solutions to massive IIA, with M$_4$ a 4d K\"ahler manifold and four Poincar\'e supercharges.
\noindent
 
\vfill
\eject

\end{titlepage}

\setcounter{footnote}{0}

\tableofcontents

\setcounter{footnote}{0}
\renewcommand{\theequation}{{\rm\thesection.\arabic{equation}}}

\section{Introduction}

\paragraph{} Two-dimensional ${\cal N}=(0,4) $ CFTs play a prominent role in the microscopic description of 5d black holes \cite{Maldacena:1997de,Vafa:1997gr,Minasian:1999qn,Castro:2008ne,Haghighat:2015ega,Couzens:2019wls}. They are also central in the description of 6d (1,0) CFTs deformed away from the conformal point. In fact, when the M5-branes are separated in an extra transverse direction  one gets a theory of interacting strings. These strings support a $(0,4)$ supersymmetric quiver gauge theory, whose elliptic genus has been shown to capture the full supersymmetric partition function of the 6d theory \cite{Haghighat:2013tka,Gadde:2015tra}.

M2-branes suspended between parallel M5-branes lead to strings on their boundaries. We refer to them as $M$-strings \cite{Haghighat:2013gba}. For M5-branes probing  $A$-type singularities, the case that will concern us in this paper, these strings are referred as $M_A$-strings  \cite{Haghighat:2013tka}. They support 2d $(0,4)$ quiver gauge theories with unitary gauge groups.
Other general configurations of $M$-strings can be obtained for M5-branes probing D-type singularities, or ``end of the space'' M9-branes. These support  quiver gauge theories involving symplectic and orthogonal gauge groups, and exceptional gauge groups, respectively \cite{Gadde:2015tra}. More general configurations can be obtained beyond the realm of M-theory, using F-theory \cite{ Lawrie:2016axq,Couzens:2017way}. In all cases, once the quiver gauge theory is specified the elliptic genus can be computed using localisation.

Explicit AdS$_3$ holographic duals to 2d (0,4) quiver gauge theories were however quite rare in the literature, with known examples reducing to intersections of D1-D5 branes \cite{Maldacena:1997re}  with KK-monopoles \cite{Kutasov:1998zh,Sugawara:1999qp,Larsen:1999dh,Okuyama:2005gq} or D9-branes \cite{Douglas:1996uz}. The recent results in \cite{Lozano:2019emq} significantly contributed to fill this gap\footnote{See the papers \cite{Kim:2005ez}-\cite{Legramandi:2019xqd} for more general AdS$_3$ solutions with different amounts of supersymmetries.}.

The geometries constructed in \cite{Lozano:2019emq} are AdS$_3\times$ S$^2\times$ M$_4$ foliations over an interval, with M$_4$ either a CY$_2$ (class I) or a 4d K\"ahler (class II) manifold. They are  solutions to massive Type IIA supergravity involving  D2-D4-D6-NS5-D8 brane configurations. They preserve small (i.e. with only one SU$(2)$ R-symmetry) ${\cal N}=(0,4)$ supersymmetries and posses an SU(2) structure. The dual CFTs of the first class were studied in \cite{Lozano:2019jza,Lozano:2019zvg}. They arise in the infrared limit of (0,4) quiver gauge theories containing two families of unitary gauge groups, $\prod_{i=1}^n \text{SU}(k_i)\times \text{SU}({\tilde k}_i)$. The gauge group SU$(k_i)$ is  associated to $k_i$ D2-branes while the gauge group SU$({\tilde k}_i)$ is associated to ${\tilde k}_i$ D6-branes, wrapped on the $\text{CY}_2$. Both D2 and D6  branes are stretched between  NS5-branes. On top of this, there are  D4 and D8 branes that provide flavour groups to both types of gauge groups, and render the field theory anomaly-free. 

The uplift of these solutions to M-theory provides explicit holographic duals to the 2d $(0, 4)$ quiver gauge theories with unitary gauge groups supported by $M_A$-strings. We will see that they are AdS$_3\times $S$^3/\mathbb{Z}_k \times$ CY$_2$ foliations over an interval that still realise small $(0,4)$ superconformal symmetry. This will be one of the main results in this paper.  Our class of solutions extends previous results in the literature, which took more restricted ansatze for the fluxes \cite{Colgain:2010wb}.  Furthermore, we are able to show that they are in one to one correspondence with 2d quiver CFTs describing $M_A$-strings. The CFTs arise as infrared fixed points of 2d field theories living on M2-branes and M-theory Kaluza-Klein monopoles (wrapped on the $\text{CY}_2$, and thus behaving effectively as 2-branes) stretched between parallel M5-branes. This set-up realises two families of unitary gauge groups, supported by flavour groups coming from extra M5-branes that render the quivers non-anomalous. Our field theories generalise quivers constructed in the literature  \cite{Gadde:2015tra}. The key ingredient is that we are able to obtain them within controlled string theory set-ups with known holographic duals. They provide examples for 2d $(0, 4)$ quiver gauge theories for which the elliptic genus can be computed. 

The contents of the paper  are distributed as follows. In section 2 we summarise the main properties of the backgrounds of the form AdS$_3\times$ S$^2\times$ CY$_2$ foliated over an interval constructed in \cite{Lozano:2019emq}. We focus our attention on compact Calabi-Yau 2-folds.  We  give a brief account of the 2d $(0,4)$ quiver CFTs  that are dual to these solutions \cite{Lozano:2019jza,Lozano:2019zvg}. In section 3 we perform the uplift of the sub-class of solutions with vanishing Romans' mass to eleven dimensions. We construct the explicit 2d quivers dual to these backgrounds and compute the central charge, both holographically and field theoretically, finding agreement in the holographic limit.  Furthermore, we show that the central charge scales linearly with the total number of $M_A$-strings of the configuration. This identifies the latter with the defining degrees of freedom of our theories, and allows us to reinterpret with generality previous results obtained in more restricted scenarios  \cite{Kutasov:1998zh}. In section 4 we construct new AdS$_3/\mathbb{Z}_k \times$ S$^3 \times$ CY$_2$ solutions in M-theory, foliated over an interval, preserving four Poincar\'e supersymmetries. We achieve this through a double analytical continuation. The new solutions are associated to M2-M5-M5' brane intersections with momentum charge, and provide a holographic description for the superconformal quantum mechanics (SCQM) that arises in the low energy limit. These SCQMs generalise quantum mechanical descriptions of M-branes in the context of M(atrix) theory studied in the literature \cite{Berkooz:1996is,Aharony:1996bh,Aharony:1997th,Aharony:1997an,Hanany:1997xc,Kachru:1998rk}. In section 5 we construct a new family of AdS$_2\times$ S$^3\times $ CY$_2\times$ I solutions to Type IIA with four Poincar\'e supercharges, upon reduction from M-theory. These solutions are associated to D0-F1-D4-D4' brane intersections. 
 We naturally extend them to backgrounds of  massive IIA supergravity, upon double analytical continuation from the solutions summarised in section \ref{AdS3S2}.
In section 6 we present our conclusions and future lines of research motivated by our results. Appendix \ref{appendix1} summarises the class I and class II solutions presented in  \cite{Lozano:2019emq}. In Appendix \ref{appendix2} we present the most general class of AdS$_3\times $S$^2 \times$ M$_4$ solutions to M-theory with (0,4) supersymmetries and SU(2) structure. Appendix \ref{appendix3} contains more general AdS$_2\times $ S$^3\times $M$_4$ solutions to massive IIA where M$_4$ is a 4d K\"ahler manifold. The geometries studied in the main body of the paper are special cases of those in the appendices. It would be interesting to understand the holographic dual to these more general backgrounds.

\section{Review of the AdS$_3\times$ S$^2\times$ CY$_2$ solutions to massive Type IIA 
}\label{AdS3S2}

In  \cite{Lozano:2019emq} the most general class of AdS$_3\times$ S$^2$ solutions to massive IIA supergravity with small (0,4) supersymmetry and SU(2) structure was presented. These solutions are foliations of AdS$_3 \times $S$^2 \times $M$_4$ over an interval, with M$_4$ either a CY$_2$ or a 4d K\"ahler manifold. The first type of solutions were referred to as class I. The second, which contain as a particular case the T-duals of the solutions found in \cite{Couzens:2017way}, were referred to as class II. The backgrounds in class I for which the symmetries of the CY$_2$ are respected by the solution constitute a particularly interesting subclass for which the full family of 2d (0,4) dual CFTs can be identified  \cite{Lozano:2019jza,Lozano:2019zvg}. This subclass of solutions is the focus of our main interest in this work. From them we will construct a general class of AdS$_3\times $S$^3/\mathbb{Z}_k\times $ CY$_2$ solutions to M-theory, to which we will associate 2d (0,4) quiver CFTs supported by $M_A$-strings.
 The uplifts of the most general solutions in class I and class II will be presented in Appendix \ref{appendix2}. Our solutions provide altogether a complete classification of AdS$_3$ solutions to M-theory with (0,4) supersymmetries and SU(2) structure.

We begin our analysis by reviewing the class I geometries constructed in  \cite{Lozano:2019emq}, with the further restriction that the symmetries of the Calabi-Yau 2-fold are respected by the full solution. This requires the Calabi-Yau to be compact, and therefore we will take it to be either $T^4$ or $K3$.
The  NS sector of this subclass of solutions reads,
\begin{equation}\label{eq:class I background}
\begin{split}
\text{d}s^2 &= \frac{u}{\sqrt{\hat{h}_4 h_8}} \left( \text{d}s^2_{\text{AdS}_3} + \frac{\hat{h}_4 h_8}{4 \hat{h}_4 h_8 + (u')^2} \text{d}s^2_{S^2} \right) + \sqrt{\frac{\hat{h}_4}{h_8}} \text{d}s^2_{\text{CY}_2} + \frac{\sqrt{\hat{h}_4 h_8}}{u} \text{d} \rho^2 \, , \\
e^{- \Phi}&= \frac{h_8^{3/4}}{2 \hat{h}_4^{1/4} \sqrt{u}} \sqrt{4 \hat{h}_4 h_8 + (u')^2} \, , \quad H_{(3)} = \frac{1}{2} \text{d} \bigg( - \rho + \frac{u u'}{4 \hat{h}_4 h_8 + (u')^2} \bigg) \wedge \widehat{\text{vol}}_{\text{S$^2$}}  \, .
\end{split}
\end{equation}
Here $\Phi$ is the dilaton, $H_{(3)}$ the NS three-form and the metric is given in string frame.  A prime denotes a derivative with respect to $\rho$.

The RR sector reads
\begin{equation}\label{eq:class I background RR}
\begin{split}
F_{(0)} &= h_8' \, , \quad F_{(2)} =  - \frac{1}{2} \Big( h_8 - \frac{h_8' u' u}{4 h_8 \hat{h}_4+(u')^2} \Big) \hat{\text{vol}}_{\text{S$^2$}} \, , \\
F_{(4)} &= -\left( \text{d} \bigg( \frac{u'u}{2 \hat{h}_4} \bigg) + 2 h_8 \text{d} \rho \right) \wedge \hat{\text{vol}}_{\text{AdS$_3$}}   - \partial_{\rho} \hat{h}_4 \hat{\text{vol}}_{\text{CY$_2$}}.
\end{split}
\end{equation}
Higher RR fluxes are related to these as $F_{(6)} = - \star F_{(4)}$, $F_{(8)} = \star F_{(2)}$, $F_{(10)} = - \star F_{(0)}$, where $\star$ is the ten-dimensional Hodge-dual operator.
Supersymmetry holds when
\begin{equation}\label{supersymmetry conditions}
u'' = 0 \,  ,
\end{equation}
which makes $u$ a linear function of $\rho$. In turn, the Bianchi identities of the fluxes impose
\begin{eqnarray}
h_8''=0\, ,~~\hat{h}_4'' =0,
\end{eqnarray}
which make $h_8$ and $\hat{h}_4$ also linear functions.
%
The particular configurations reviewed above are independent of the  CY$_2$-fold coordinates and $\widehat{h}_4$ has support on the $\rho$ coordinate only.
The supersymmetry and Bianchi identities are  satisfied for $u$, $h_8$, $\hat{h}_4$ arbitrary linear functions in $\rho$. This is the  above mentioned restriction we adopt with respect to \cite{Lozano:2019emq}.
We shall keep this restriction hereafter, with the exception of the material in the appendices.
\\
The magnetic components of the Page fluxes ${\hat F}=F\wedge e^{-B_{(2)}}$ are given by
\begin{eqnarray}
&&{\hat F}_{(0)} = h_8'\\ \label{F0}
&&{\hat F}_{(2)}=-\frac12 \Bigl(h_8-(\rho-2 \pi j) h_8'\Bigr)\hat{\text{vol}}_{\text{S$^2$}}\\
&&{\hat F}_{(4)}=- \hat{h}_4'  \hat{\text{vol}}_{\text{CY$_2$}}\\
&&{\hat F}_{(6)}=\frac12  \Bigl(\hat{h}_4-(\rho-2 \pi j) \hat{h}_4'\Bigr) \hat{\text{vol}}_{\text{CY$_2$}}\wedge \hat{\text{vol}}_{\text{S$^2$}}, \label{F6}
\end{eqnarray}
where we have included large gauge transformations in $B_{(2)}$ of parameter $j$, such that
\begin{equation}
B_{(2)}=\frac12 \bigg( 2 \pi j -\rho +\frac{u u'}{4 \hat{h}_4 h_8 + (u')^2} \bigg) \wedge \hat{\text{vol}}_{\text{S$^2$}}\, . 
\end{equation}

\subsection{Brief description of the 2d dual CFTs}

\begin{figure}
\centering
\includegraphics[scale=0.5]{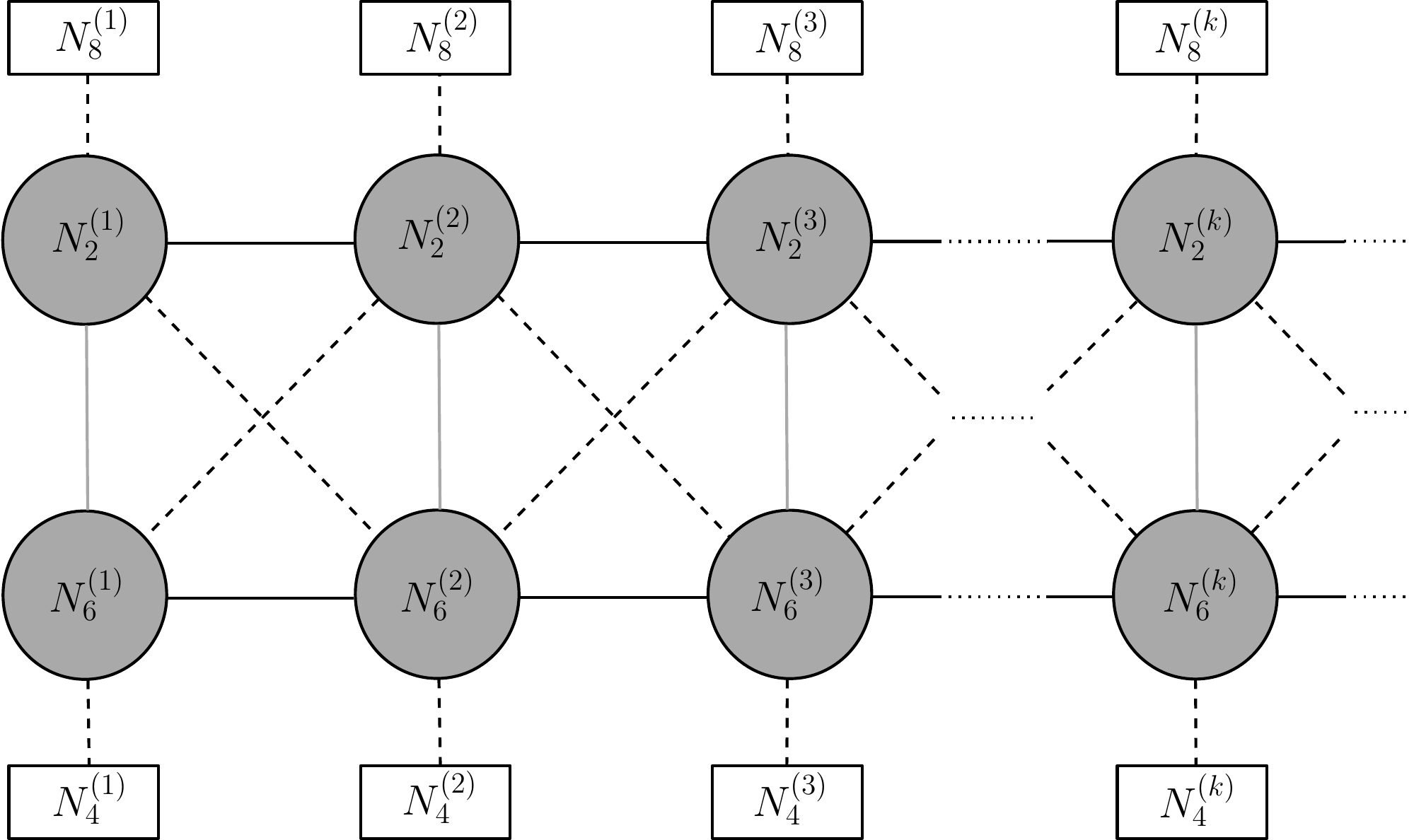}
\caption{Generic quiver field theory whose IR is holographic dual to the solutions reviewed in this section. The solid black line represents a (4,4) hypermultiplet, the grey line a (0,4) hypermultiplet and the dashed line a (0,2) Fermi multiplet.  The degrees of freedom at each node are (4,4) vector multiplets.}
\label{general-quiver}
\end{figure}

\paragraph{} Associated to the Page fluxes there is a D2-D4-D6-D8-NS5 brane system, depicted in Table \ref{branesetupAdS3IIA}. 
  \begin{table}[ht]
	\begin{center}
		\begin{tabular}{| l | c | c | c | c| c | c| c | c| c | c |}
			\hline		    
			& 0 & 1 & 2 & 3 & 4 & 5 & 6 & 7 & 8 & 9 \\ \hline
			D2 & x & x & &  &  &  & x  &   &   &   \\ \hline
			D4 & x & x &  &  &  &   &  & x & x & x  \\ \hline
			D6 & x & x & x & x & x & x & x  &   &   &   \\ \hline
			D8 & x & x &x  & x & x &  x &  & x & x & x  \\ \hline
			NS5 & x & x &x  & x & x & x  &   &   &  &  \\ \hline
		\end{tabular} 
	\end{center}
	\caption{$\frac18$-BPS brane intersection underlying the AdS$_3$ solutions. $(x^0,x^1)$ are the directions where the 2d CFT lives, $(x^2, \dots, x^5)$ span the CY$_2$, on which the D6, the D8 and the NS5 branes are wrapped, $x^6$ is the direction along the $\rho$-interval, and $(x^7,x^8,x^9)$ are the transverse directions on which the SO$(3)_R$ symmetry is realised.}   
	\label{branesetupAdS3IIA}
\end{table} 
The 2d CFTs that live on these brane set-ups were analysed in  \cite{Lozano:2019jza,Lozano:2019zvg}, to which the reader is referred for more details. They are described by 
 $(0,4)$ superconformal quivers with gauge groups associated to stacks of D2 and D6 branes (the latter wrapped on the CY$_2$ manifold),  both stretched between NS5 branes. Being the extension of the D2 and D6 branes finite in the $\rho$ direction, the field theory living on their intersection is effectively two dimensional 
 %
 at low energies. These quivers are rendered non-anomalous with adequate flavour groups at each node, coming from D4 and D8 branes. Figure \ref{general-quiver} illustrates their general structure. 
The quivers can be divided into two long linear quivers consisting on $(4,4)$ gauge groups connected horizontally by $(4,4)$ bifundamental hypermultiplets, coupled through $(0,4)$ hypermultiplets (vertically) and $(0,2)$ Fermi multiplets (in the diagonals). The flavour degrees of freedom couple through $(0,2)$ Fermi multiplets with its corresponding gauge node. These couplings render the quiver $(0,4)$ supersymmetric.

Let us  see how the cancellation of gauge anomalies works. For a given SU$(N_2^{(i)})$ gauge group we are concerned with the contributions to the anomaly coming from the $(0,4)$ hypermultiplets that connect it to the SU$(N_6^{(i)})$ gauge node and with the various $(0,2)$ Fermi multiplets. The $(0,4)$ hypermultiplets in the bifundamental representation are composed of two $(0,2)$ chiral multiplets, which contribute to the gauge anomaly a factor of $1$. In turn, the $(0,2)$ Fermi multiplets in the fundamental or bifundamental representations contribute a factor of $-\frac12$.  Putting these together, we have that for a SU$(N_2^{(i)})$ gauge group the gauge anomaly cancellation condition is
\begin{equation}
2 N_6^{(i)}=N_6^{(i-1)}+N_6^{(i+1)}+N_8^{(i)}\, .
\end{equation}
This becomes
\begin{equation}
\label{fortheN6}
2 N_2^{(i)}=N_2^{(i-1)}+N_2^{(i+1)}+N_4^{(i)}\, ,
\end{equation}
for SU$(N_6^{(i)})$ gauge groups.  The reader is referred to  \cite{Lozano:2019zvg} for more details. 
\\
In the next section we study the M-theory lift of the backgrounds in (\ref{eq:class I background})-(\ref{eq:class I background RR}).

\section{New AdS$_3\times$ S$^3/\mathbb{Z}_k\times $ CY$_2$ solutions in M-theory}\label{Mtheoryuplift}

Let us consider the uplift to eleven dimensions of the solutions
discussed in the previous section. To perform this lift we need $F_{(0)}=0$, which according to (\ref{eq:class I background RR}) imposes the function $h_8$ to be a constant. 
Thus, the IIA brane configuration that we lift consists on intersecting D2-D4-D6-NS5 branes. The restriction to vanishing Romans' mass implies that the number of D6-branes ($N_6$) must remain constant between all pairs of NS5-branes. 
In the lift to eleven dimensions this number becomes a modding parameter of the geometry, associated with KK-monopole charge.

 Once this lift  is performed, we obtain a class of AdS$_3\times $S$^3/\mathbb{Z}_k\times$  CY$_2$  solutions to M-theory foliated over an interval. They preserve ${\cal N}=(0,4)$ 
 supersymmetry. These solutions read
\begin{eqnarray}
\label{M-theory}
\text{d} s_{11}^2&=&\Delta\left(\frac{u}{\sqrt{\hat{h}_4 h_8}} \text{d}s_{\text{AdS}_3}^2+\sqrt{\frac{\hat{h}_4}{h_8}} \text{d}s_{\text{CY}_2}^2+\frac{\sqrt{\hat{h}_4 h_8}}{u} \text{d} \rho^2
\right)+\frac{h_8^2}{\Delta^2} \text{d} s^2_{\text{S}^3/\mathds{Z}_k}\label{Mmetric} \, ,\\
G_{(4)}&=&-\text{d}\left(\frac{uu'}{2\hat{h}_4}+2\rho h_8\right)\wedge \hat{\text{vol}}_{\text{AdS}_3}+2h_8\;\text{d}\left(-\rho+\frac{u u'}{4\hat{h}_4h_8+u'^2}\right)\wedge \hat{\text{vol}}_{\text{S}^3/\mathds{Z}_k}\nonumber \label{M4flux} \\
&&-\partial_{\rho}\hat{h}_4\;\hat{\text{vol}}_{\text{CY}_2} \, , \\
 \Delta &=&\frac{h_8^{1/2}(4\hat{h}_4h_8+u'^2)^{1/3}}{2^{2/3}\hat{h}_4^{1/6}u^{1/3}}\, ,\nonumber
\end{eqnarray}
where $k=h_8=N_6$. The quotiented 3-sphere is written as an $S^1_z$ Hopf fibration over an S$^2$,
\begin{eqnarray}
\text{d} s^2_{\text{S}^3/\mathds{Z}_k}=\frac{1}{4}\left[\left(\frac{ \text{d} z}{k}+\eta\right)^2+\text{d} s^2_{\text{S}^2}\right]\qquad\text{with}\qquad \text{d} \eta=\hat{\text{vol}}_{\text{S}^2}.
\label{saza}\end{eqnarray}
In the previous solutions the symmetries $\text{SL}(2, \mathbb{R}) \times \text{SL}(2, \mathbb{R})$ and $\text{SU}(2)$ are realised geometrically on the AdS$_3$ and the quotiented 3-sphere, respectively.
%

\begin{figure}
\centering
\includegraphics[scale=0.6]{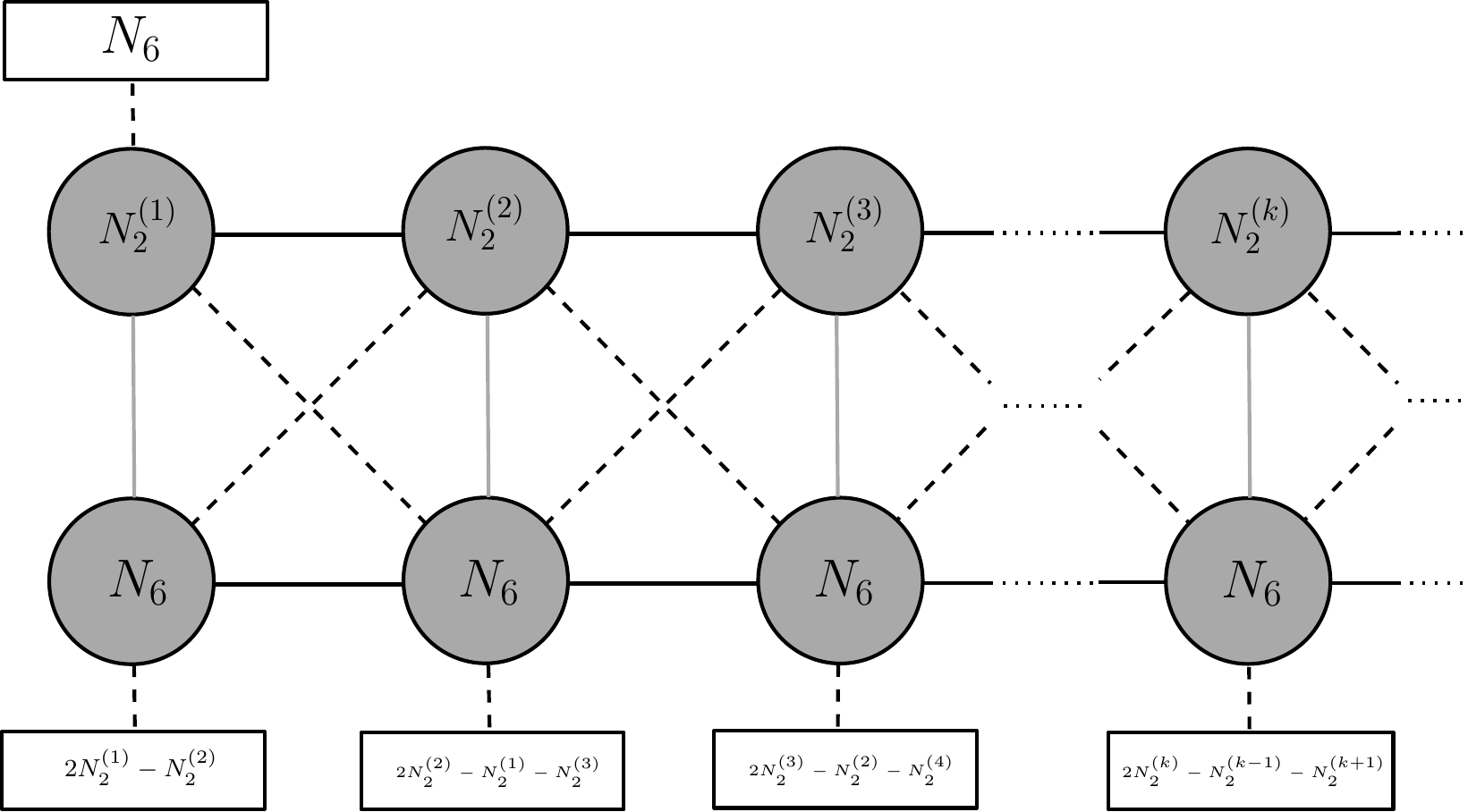}
\caption{Generic quiver field theories dual to the AdS$_3$ solutions with vanishing Romans' mass.}
\label{generalM}
\end{figure}

The dual quivers associated to this class of solutions are depicted in 
Figure \ref{generalM}. The gauge anomaly is automatically cancelled for the SU$(N_2^{(i)})$ gauge groups, once an extra SU$(N_6)$ flavour group is added to the first node, while for the SU$(N_6)$ gauge groups the condition (\ref{fortheN6}) has been enforced.
In what follows, we concentrate on the  backgrounds in (\ref{Mmetric})-(\ref{M4flux}). In appendix \ref{appendix2} we discuss the lift to eleven dimensions of the more general backgrounds constructed in \cite{Lozano:2019emq}.
%
   %
  %
%
\subsection{Brane set-up} \label{brane-set-up}

\paragraph{} 
In the new class of solutions given by (\ref{Mmetric})-(\ref{M4flux}), the number of Type IIA D6-branes became the orbifold parameter in S$^3/\mathds{Z}_k$, $k=N_6=h_8$, and thus  corresponds to KK-monopole charge. The D2-branes became M2-branes. Their charge in the interval $\rho\in [2\pi j,2\pi (j+1)]$ is obtained by integrating the Page flux ${\hat G}_{(7)}=G_{(7)} - G_{(4)} \wedge C_{(3)}$. The component of $\hat{G}_7$ relevant to calculate the number of M2 branes is given by
\begin{equation}
\label{G7}
{\hat G}_{(7)} = 2h_8 \Bigl(\hat{h}_4 - (\rho - 2 \pi j) \hat{h}_4'\Bigr)  \hat{\text{vol}}_{\text{S}^3/\mathds{Z}_k} \wedge \hat{\text{\text{vol}}}_{\text{CY}_2}\, .
\end{equation}

The D4-branes of the Type IIA solution became M5-branes. Their presence is captured by non-trivial flux of $G_4$ through the CY$_2$. Finally, the NS5 branes became M5'-branes, whose charge is given by a non-trivial flux of $G_4$ through the $(\rho, S^3/\mathds{Z}_k)$ cycle. Therefore, the D2-D4-D6-NS5 branes underlying the Type IIA solutions become M2-M5-KK-M5' branes, intersecting as shown in Table \ref{branesetupAdS3M}. The KK-monopoles (wrapped on the $\text{CY}_2$) and the M2 branes are stretched between parallel M5'-branes and there are extra M5-branes providing for flavour groups. This describes $M_A$-strings, supplemented with extra M5-branes.
 \begin{table}[ht]
	\begin{center}
		\begin{tabular}{| l | c | c | c | c| c | c| c | c| c | c | c|}
			\hline		    
			& 0 & 1 & 2 & 3 & 4 & 5 & 6 & 7 & 8 & 9 & 10 \\ \hline
			M2 & x & x & &  &  &  & x  &   &   &  & \\ \hline
			M5 & x & x &  &  &  &   &  & x & x & x & x \\ \hline
			KK & x & x & x & x & x & x & x  &   &   &  & z  \\ \hline
			M5' & x & x &x  & x & x & x  &   &   &  &  &  \\ \hline
		\end{tabular} 
	\end{center}
	\caption{$\frac18$-BPS brane intersection underlying the AdS$_3\times$ S$^3/\mathbb{Z}_k$ solutions in M-theory.  The directions $(x^0,x^1)$ are those where the 2d dual CFT lives, $(x^2, \dots, x^5)$ span the CY$_2$, $x^6$ is the `field space' direction and $(x^7,x^8,x^9)$ are the transverse directions on which the SO(3)$_R$ symmetry is realised. Finally, $x^{10}$ is the extra eleventh direction, which spans the $S^1/\mathbb{Z}_k\subset S^3/\mathbb{Z}_k$ and plays the role of Taub-NUT direction of the KK-monopole.}  
	\label{branesetupAdS3M}
\end{table} 
The corresponding dual quivers are the ones depicted in Figure \ref{generalM}, with upper row nodes associated to M2-branes and lower row nodes to KK-monopoles. The M5-branes provide for extra flavour groups that render the quivers non-anomalous (and the supergravity equations of motion satisfied).

Our new solutions in M-theory (\ref{Mmetric})-(\ref{M4flux}), provide for explicit AdS$_3$ geometries that can be used to study these quivers holographically. It would be interesting to see these geometries emerging in the near horizon limit of intersecting M2-M5-MKK-M5' brane systems. This is currently under investigation \cite{FLP}. 



Note that when $u'=0$ the M5-branes wrapped on $AdS_3\times S^3/\mathbb{Z}_k$ support self-dual strings on their worldvolumes, coupled to the (self-dual) 3-form field
\begin{equation}
C_{(3)} = - 2 \rho \, h_8 ( \hat{{\rm vol}}_{AdS_3} + \hat{{\rm vol}}_{S^3/\mathbb{Z}_k})\, .
\end{equation}
They arise from M2-branes ending on the M5-branes. These solutions provide then for fully backreacted near horizon geometries for OM theory \cite{Berman:2001fs}, the theory conjectured to be the UV completion of the (2,0) theory with constant background 3-form field living on the M5-branes \cite{Gopakumar:2000ep}. In our explicit set-up the 3-form depends on the positions of the M5-branes in the $\rho$-direction. Extra intersecting M5'-branes and KK-monopoles further reduce the supersymmetries by a half. 
  
\begin{figure}
\centering
\includegraphics[scale=0.8]{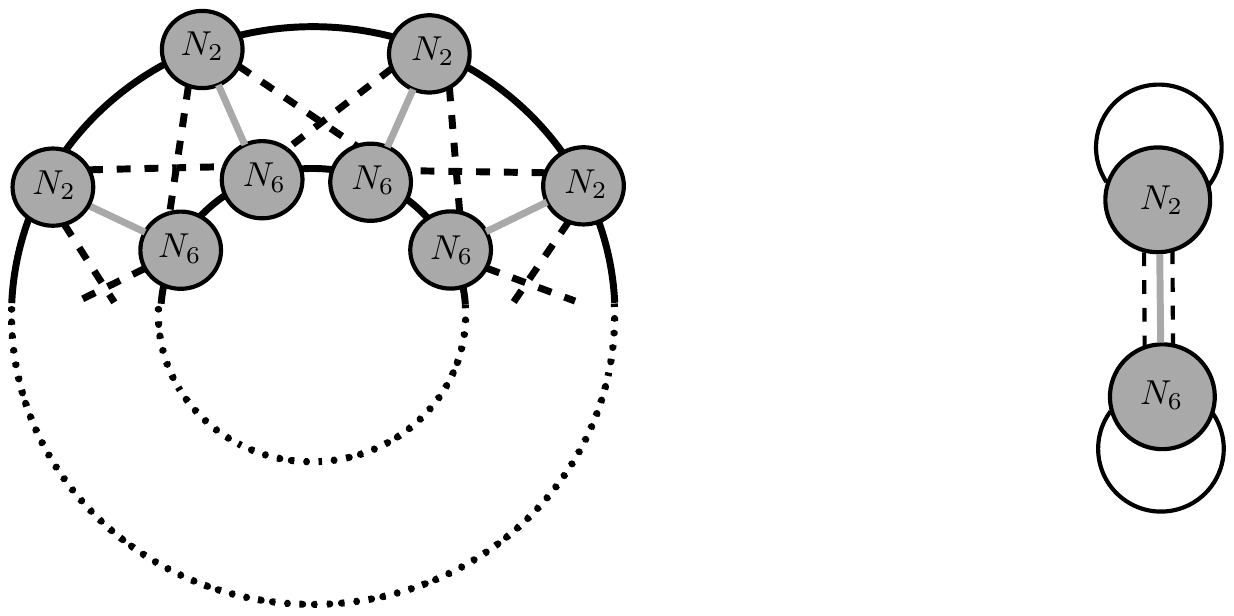}
\caption{Left: Generic quiver field theory whose IR limit is holographic dual to the $AdS_3$ solutions with $I=S^1$. Right: Quiver field theory for $M=1$. In the right quiver, the (0,4) hypermultiplets combine with two (0,2) Fermi multiplets to produce (4,4) hypermultiplets. Supersymmetry is thus  enhanced to (4,4).}
\label{h4constant}
\end{figure}

An interesting particular case contained in our class of solutions is when $\mathcal{I} = S^1$, for which $\hat{h}_4'=u'=0$. This case was discussed in \cite{Haghighat:2013tka} (see also \cite{Kim:2015gha}). In this case the background  \cite{Boonstra:1998yu} is the uplift of the T-dual of the $AdS_3\times S^3/\mathbb{Z}_M\times \text{CY}_2$ geometry that describes D1-D5 branes in a $A_{M-1}$ singularity, introduced by $M$ KK-monopoles. The IIB KK-monopoles become the M5'-branes in M-theory, with their Taub-NUT charge provided by the Type IIB D5-branes. The associated 2d quivers are those on the left of Figure \ref{h4constant}. When $M=1$ supersymmetry is enhanced to (4,4), the brane system becomes a M2-M5' brane intersection and the associated quiver becomes the one on the right.

\begin{figure}
\centering
\includegraphics[scale=0.7]{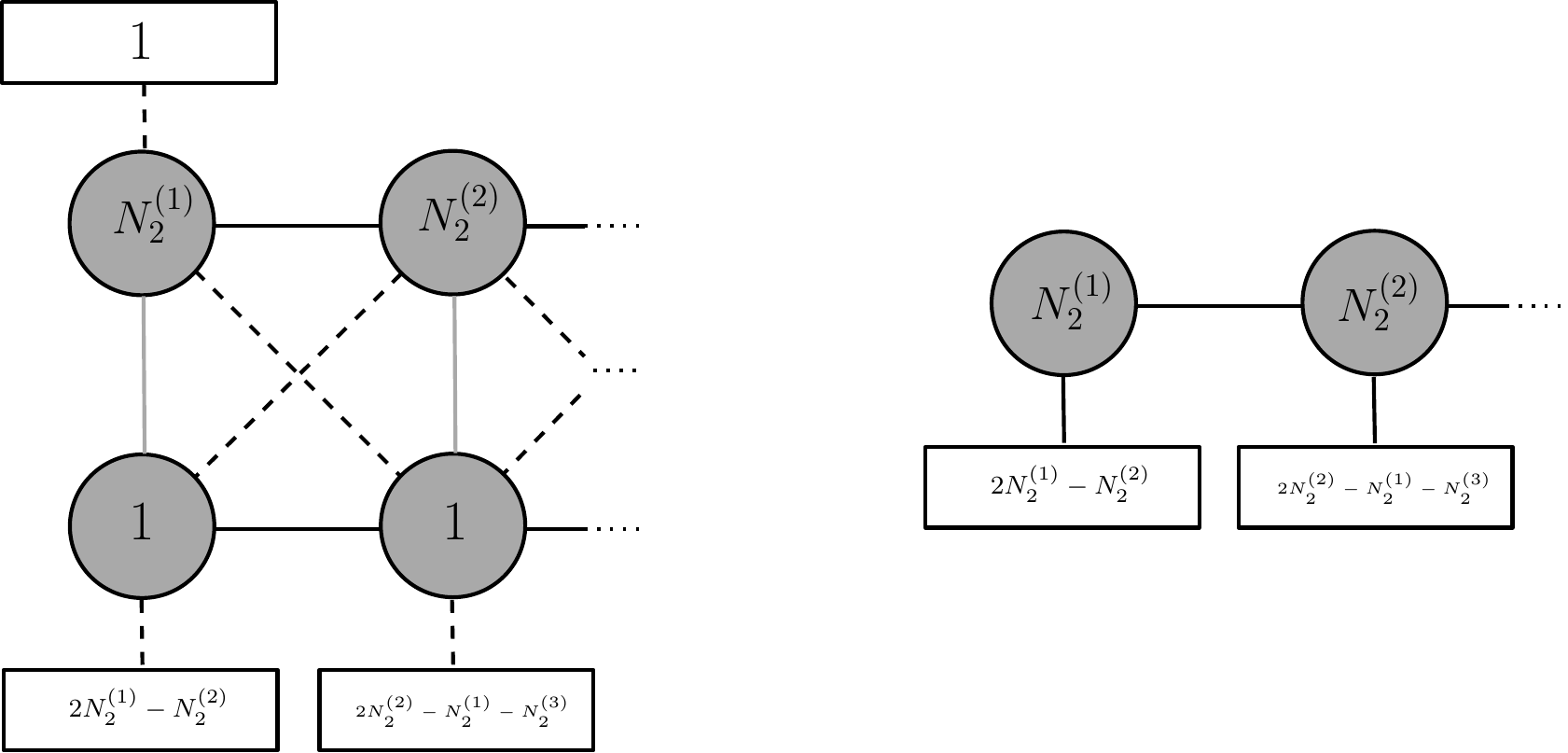}
\caption{Left: 2d (0,4) quiver CFT dual to the AdS$_3\times$ S$^3\times $CY$_2\times$ I solution. Right: 4d $\mathcal{N}=2$ quiver CFT with flavours.}
\label{decons}
\end{figure}

Another interesting case is when $k=1$ and  there is just one KK-monopole stretched between the M5'-branes. The resulting quivers are depicted in Figure \ref{decons} (left).
In M-theory one KK-monopole is equivalent to no modding, and therefore the brane system reduces to the M2-M5-M5' brane intersection included in Table \ref{branesetupAdS3M}. This intersection is still 1/8-BPS. These brane intersections might play a role as brane set-ups where 2d defect 
CFTs could be realised, in connection with the phenomenon of deconstruction \cite{ArkaniHamed:2001ca}. Indeed, our quivers generalise (by the inclusion of flavours) the 2d defect CFTs living in D3-D3'-KK intersections studied in  \cite{Constable:2002vt}, which deconstruct 4d $\mathcal{N}=2$ CFTs living in M5-brane intersections. One might expect that the 2d quivers depicted on the left of Figure \ref{decons}
could emerge through a similar mechanism as the one described in \cite{Constable:2002vt},  by coupling the 4d $\mathcal{N}=2$ CFT depicted on the right with an Abelian field theory. It would be interesting to explore this possibility.

\subsection{Central charge}

\begin{figure}
\centering
\includegraphics[scale=0.7]{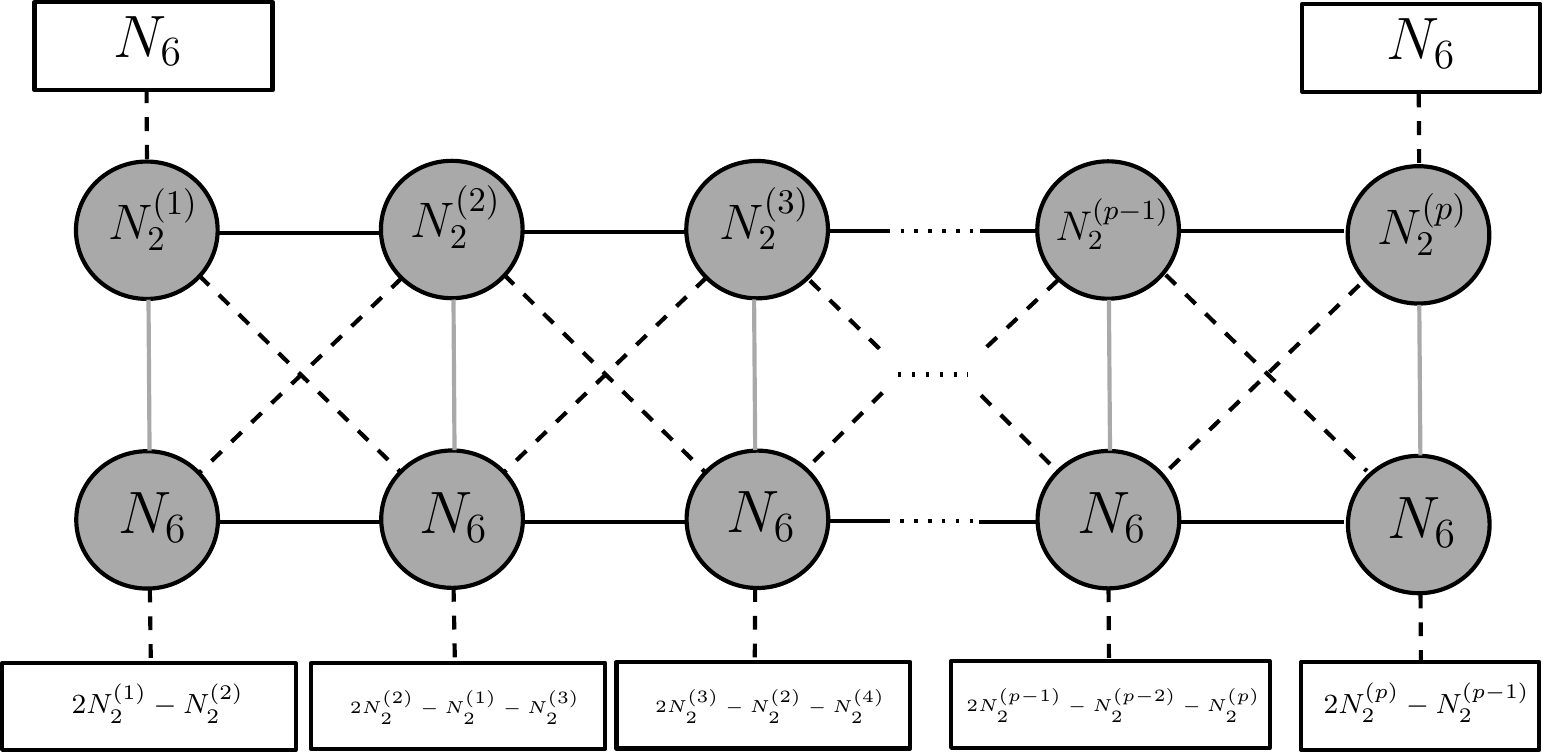}
\caption{Completed quiver field theories whose IR limits are holographic duals to the AdS$_3\times$ S$^3/\mathbb{Z}_k$ solutions in M-theory. $N_2^{(j)}$ refer to M2-brane charges and $N_6=k$ to the constant, KK-monopole charge. M5-branes provide for the $2N_2^{(i)}-N_2^{(i-1)}-N_2^{i+1)}$ flavour groups that render the quiver non-anomalous.}
\label{completed}
\end{figure}

\paragraph{} In this section we compute the  (right moving) central charge of the CFTs dual to our solutions. We consider generic quivers such as the ones depicted in Figure \ref{generalM}, that we terminate by adding adequate flavour groups, rendering the quiver non-anomalous, with  large but finite length (see 
\cite{Lozano:2019zvg} for more details). One possibility is the completed quiver depicted in Figure \ref{completed}. 
The corresponding functions  $h_8$ and  $\hat{h}_4(\rho)=\Upsilon h_4(\rho)$  are given by,
 \begin{equation} \label{profileh8exampleII}
h_8= N_6,~~ 0\leq \rho\leq 2 \pi(P+1).\nonumber
\end{equation}
\begin{equation}\label{profileh4sp}
\hat{h}_4(\rho)\!=\Upsilon
                    \!\!\left\{ \begin{array}{ccrcl}
                       \frac{\beta_0 }{2\pi}
                       \rho & 0\leq \rho\leq 2\pi \\                
                                     \alpha_j\! +\! \frac{\beta_j}{2\pi}(\rho-2\pi j) &~~ 2\pi j\leq \rho \leq 2\pi(j+1), \\
                     \alpha_P-  \frac{\alpha_P}{2\pi}(\rho-2\pi P) & 2\pi P\leq \rho \leq 2\pi(P+1),
                                             \end{array}
\right.
\end{equation}
with $\alpha_j=\sum_{r=0}^{j-1}\beta_r$ and $j=1,\dots, P-1$.
We have  $\hat{h}_4(0)=\hat{h}_4(2\pi(P+1))=0$. 
 At these values of $\rho$, the asymptotic analysis indicates the presence of M5-branes extended on AdS$_3\times$ S$^3/\mathbb{Z}_k$ and the space terminates. In what follows, we choose $\Upsilon$ such that $\Upsilon {\rm vol}_{\text{CY}_2}=16\pi^4 $.

The numbers of M2 and M5 branes at each $2\pi j\leq \rho \leq 2\pi(j+1)$ interval, with $j=1,\dots, P$,  are given by
\begin{equation}
\label{numberM2}
N_2^{(j)}=\frac{1}{(2\pi)^6}\int_{\text{CY}_2\times S^3/\mathbb{Z}_k} {\hat G}_{(7)}=\frac{2}{(2\pi)^6}\Bigl(\hat{h}_4-(\rho-2\pi j)\hat{h}_4'\Bigr){\rm vol}_{\text{CY}_2}{\rm vol}_{S^3}=\alpha_j
\end{equation}
and
\begin{equation}
N_5^{(j)}=\frac{1}{(2\pi)^3}\int_{\text{CY}_2}G_{(4)}=\left\{ \begin{array}{ccrcl}
                                                                    \beta_j &  2\pi j\leq \rho \leq 2\pi(j+1); \, \,\,  j=0,\dots, P-1 \\
                                                                     -\alpha_P & 2\pi P\leq \rho \leq 2\pi(P+1).
                                                                     \end{array}
\right.
\end{equation}
Notice that $\beta_{j-1}-\beta_j=2N_2^{(j)}-N_2^{(j-1)}-N_2^{(j+1)}$ is the number of flavours at each $2\pi j\leq \rho \leq 2\pi(j+1)$ interval, with $j=1,\dots,P-1$, and there are extra $\alpha_P+\beta_{P-1}= 2N_2^{(P)}-N_2^{(P-1)}$ flavours at the $2\pi P\leq \rho \leq 2\pi(P+1)$ interval, as depicted in Figure \ref{completed}.

 We proceed now with the computation of the holographic central charge. The interesting result we shall obtain is that
 the central charge is proportional to the total number of $M_A$-strings and is also related to the action of the $M_5'$-branes. 
 \\
 The central charge being directly proportional to the number of $M_A$-strings indicates that the fundamental degrees of freedom of this theory should be understood as $M_A$-strings. On the other hand, the relation between the action of $M_5'$-branes and the central charge indicates that these branes (that provide a boundary condition for the membranes to end) capture on their world-volumes the dynamics of the lower dimensional branes. This is a non-trivial fact, already encountered in \cite{Dibitetto:2019nyz}. It would be of interest to reproduce it in other holographic systems to fully understand its origin.

The right-moving central charge of the AdS$_3\times $S$^2\times$ CY$_2$ solutions constructed in \cite{Lozano:2019emq} was computed in \cite{Lozano:2019zvg}. The expression found there remains valid upon uplift to eleven dimensions. In terms of the  ten dimensional Newton's constant we have,
\begin{equation}
\label{chol}
c=\frac{3\pi}{2G_N}{\rm vol}_{\text{CY}_2} \int \text{d} \rho \,h_8 \hat{h}_4.
\end{equation}
This gives, for the functions $\hat{h}_4$ and $h_8$ displayed above
\begin{equation}
c=\frac{3}{\pi}h_8\int_0^{2\pi(P+1)} \text{d} \rho \, h_4 = 6 h_8 \sum_{j=1}^P\alpha_j =6 h_8 \sum_{j=1}^P N_2^{(j)}=6 k N_2= 6\, N_{M_A}\, ,
\end{equation}
where $N_{M_A}$ stands for the total number of $M_A$-strings in the configuration, taking into account the orbifolding by $\mathbb{Z}_k$. 
This result emphasises the fact that the $M_A$-strings holographically capture the degrees of freedom of the  conformal field theory. This suggests that the $M_A$-strings actually are the degrees of freedom of the strongly coupled conformal field theory.

We show next that this result can be reproduced from the action describing the M5'-branes of the configuration, where the M2-branes end, realising the $M_A$-strings introduced in \cite{Haghighat:2013gba,Haghighat:2013tka}. 

The M5'-branes on which the M2-branes end, span the $(t,x^1,\text{CY}_2)$ directions of the geometry, and are positioned along the $\rho$-interval at $\rho=2\pi j$. Their worldvolume effective action is given by
\begin{equation}
S_{M5'^{(j)}}=T_{M5'}\int \text{d}^6\xi\sqrt{{\rm det}g}=\frac14 T_{M5'}{\rm vol}_{\text{CY}_2}\int \Bigl(\hat{h}_4 h_8 +\frac14 u'^2\Bigr) \cosh{r} \, \text{d} t \text{d} x^1\, .
\end{equation}
For an M5'-brane located at $\rho=2\pi j$ and $r=0$ this becomes
\begin{equation}
S_{M5'^{(j)}}=\frac{1}{4 (2\pi)^4} {\rm vol}_{\text{CY}_2} {\rm vol}_{\mathbb{R}}\, \Bigl( \hat{h}_4(2\pi j) h_8+\frac14 u'^2\Bigr) = \frac{1}{4}  {\rm vol}_{\mathbb{R}}\, \Bigl( \alpha_j  h_8+\frac{u'^2}{4\Upsilon} \Bigr).
\end{equation}
Summing the contributions of all M5'-branes we have, to leading order in $P$,
\begin{equation}
S_{M5'}=\sum_{j=1}^P S_{M5'^{(j)}}\sim h_8 \sum_{j=1}^P \alpha_j\, =N_{M_A}\, .
\end{equation}
This reproduces the scaling of the central charge to leading order within the context of the M5'-branes effective action.

Our discussion in the previous subsection about the interpretation of the $M_A$-strings as self-dual strings when $u'=0$ suggests that we should also be able to reproduce the scaling of the central charge from the M5-branes effective action, where the M2-branes realise self-dual strings. However,  to check this  we would need an action for non-Abelian M5-branes.

\subsubsection{Field theory calculation}

Finally, we check for consistency that the previous central charge coincides with the field theory result for long quivers with large ranks--the regime in which we can trust the supergravity solutions (see \cite{Lozano:2019zvg}). At the conformal point the central charge is related to the two point U(1)$_R$ current correlation function (see for example \cite{Putrov:2015jpa}), such that
\begin{equation}
c=6 (n_{hyp}-n_{vec})\, ,
\end{equation}
where $n_{hyp}$ is the number of $\mathcal{N}=(0,4)$ hypermultiplets and $n_{vec}$ the number of 
 $\mathcal{N}=(0,4)$ vector multiplets of the quiver in the UV description. 
 
 For the quivers considered in Figure \ref{completed}, we find
 \begin{equation}
 n_{hyp}=\sum_{j=1}^{P-1} N_2^{(j)}N_2^{(j+1)}+(P-1)N_6^2+N_6\sum_{j=1}^P N_2^{(j)}\, ,
 \end{equation}
 and
 \begin{equation}
 n_{vec}=\sum_{j=1}^P \Bigl((N_2^{(j)})^2-1\Bigr)+P(N_6^2-1)\, .
\end{equation}
After defining  $N_2=\sum_{j=1}^P N_2^{(j)}$, this gives for the central charge
\begin{equation}
c=6 (n_{hyp}-n_{vec})=6\Bigl(N_6N_2+\sum_{j=1}^{P-1}N_2^{(j)}N_2^{(j+1)}-\sum_{j=1}^P(N_2^{(j)})^2-N_6^2+2P\Bigr)\, .
\end{equation}
Now, the net  contribution to this expression of the term $\left(\sum_{j=1}^{P-1}N_2^{(j)}N_2^{(j+1)}-\sum_{j=1}^P(N_2^{(j)})^2\right)$ is subleading when  compared with the contribution of $N_6N_2$. This hierarchy occurs when the number of gauge nodes is large (for long quivers). As a consequence, to leading order in the number of nodes $P$, we find
\begin{equation}
c=6N_6N_2+ \mathcal{O} (P).\label{ppoo}
\end{equation}
%

\begin{figure}
\centering
\includegraphics[scale=0.8]{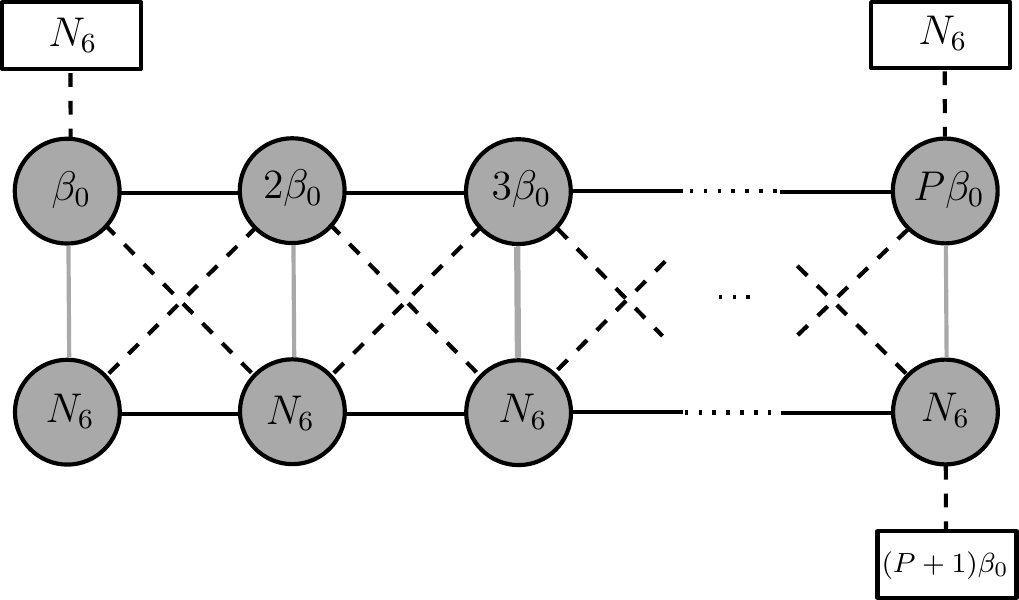}
\caption{2d (0,4) quiver CFT with gauge groups of linearly increasing ranks.}
\label{exampleCarlos}
\end{figure}

The only situation in which the two competing terms  above scale similarly in $P$, is when there are no intermediate flavour groups, {\it i.e.} when
$N_2^{(j)}=j \beta_0$ for $i=1,\dots,P$. 
This particular situation corresponds to the quiver with gauge groups of linearly increasing ranks, depicted in Figure \ref{exampleCarlos}.  
Next we  show that the contribution of 
$\left(\sum_{j=1}^{P-1}N_2^{(j)}N_2^{(j+1)}-\sum_{j=1}^P(N_2^{(j)})^2\right)$ is indeed subleading with respect to that of  $ N_6\sum_{j=1}^P N_2^{(j)}$.  To do this, we should  impose that $N_6$ is much bigger than $\beta_0$. This is required to have a trustable supergravity background--see the analysis in section 4.4 of \cite{Lozano:2019zvg}.
The central charge then reads
\begin{equation}
c=6\bigg[N_6N_2+\beta_0^2 \Bigl(\sum_{j=1}^{P-1}j(j+1)-\sum_{j=1}^P j^2\Bigr)-N_6^2+2P\bigg].
\end{equation}
Keeping in mind that $N_2=\beta_0\sum_{j=1}^P j=\beta_0\frac{P(P+1)}{2}$ we get
\begin{equation}
c=3\beta_0 (N_6-\beta_0) P^2 + \mathcal{O}(P)\sim 3\beta_0 N_6 P^2\, .\label{mmzz}
\end{equation}
We used  that $N_6$ must be much larger than $\beta_0$ for the supergravity background to be trustable. 

One can now easily check that this is in agreement with the holographic calculation.
In fact, the function $\hat{h}_4$ representing the quiver with linearly increasing ranks is 
  \begin{equation} \label{profileh4final}
\hat{h}_4(\rho)=\Upsilon h_4(\rho)
                    =\Upsilon \left\{ \begin{array}{ccrcl}
                       \frac{\beta_0 }{2\pi}
                       \rho & 0\leq \rho\leq 2\pi P\\
                        \frac{\beta_0 P}{2\pi}(2\pi (P+1)-\rho) & 2\pi P\leq \rho \leq 2\pi(P+1),
                                             \end{array}
\right.
\end{equation}
with $h_8=N_6$. Using our expression for the central charge (\ref{chol}) we find,
\begin{eqnarray}
& & c\!=\frac{3}{\pi} N_6 \left(\!\int_0^{2\pi P} \frac{\beta_0}{2\pi}\rho d\rho  +\int_{2\pi P}^{2\pi(P+1) } \frac{\beta_0}{2\pi}(2\pi(P+1)-\rho)d\rho\right)\!=\!3 N_6 \beta_0 P^2(1+ \frac{1}{P}),
\end{eqnarray}
in coincidence with eq.(\ref{mmzz}) when long quivers are considered.

Any other situation with intermediate (many, but  sparse enough) flavour groups will work along similar lines, showing the validity of eq.(\ref{ppoo}). This shows that the holographic calculation and the field theoretical one coincide for long quivers with large enough ranks. 

This closes our analysis of the backgrounds in equations (\ref{Mmetric})-(\ref{M4flux}). In the next section we present a new branch of AdS$_3$ solutions in M-theory.

\section{Double analytic continuation}

\paragraph{} A double analytic continuation in the AdS$_3 \times$ S$^3/\mathbb{Z}_k \times $ CY$_2\times $ I solutions presented in 
(\ref{Mmetric})-(\ref{M4flux}),  gives rise to a second class of solutions in which the AdS$_3$ subspace is quotiented instead of the S$^3$. These solutions preserve the same amount of supersymmetries.
 The KK-monopoles turn into M0-branes, or waves, with the Taub-NUT direction of the KK-monopoles turning into the direction of propagation of the waves. The solutions are then  associated to the M0-M2-M5-M5' brane intersections depicted in Table \ref{branesetupAdS3Mq}.  
 \begin{table}[ht]
	\begin{center}
		\begin{tabular}{| l | c | c | c | c| c | c| c | c| c | c | c|}
			\hline		    
			& 0 & 1 & 2 & 3 & 4 & 5 & 6 & 7 & 8 & 9 & 10 \\ \hline
			M0 & x & z &  &  &  &  &   &   &   &  &   \\ \hline
			M2 & x & x & &  &  &  & x  &   &   &  & \\ \hline
			M5 & x & x &  &  &  &   &  & x & x & x & x \\ \hline
			M5' & x & x &x  & x & x & x  &   &   &  &  &  \\ \hline
		\end{tabular} 
	\end{center}
	\caption{$\frac18$-BPS brane intersection underlying the AdS$_3/\mathbb{Z}_k\times$ S$^3$ solutions in M-theory. $x^1$ is the direction of propagation of the wave. $(x^2, \dots, x^5)$ span the CY$_2$, $x^6$ is the direction along the $\rho$-interval, $(x^7,x^8,x^9,x^{10})$ are the transverse directions on which the SO(4) symmetry is realised. The presence of the wave renders the dual CFT one-dimensional.}  
	\label{branesetupAdS3Mq}
\end{table}
%
The double analytic continuation of the background given in \eqref{M-theory} works as follows. The AdS$_3$ and S$^3$ factors can be swapped as
\begin{equation}
AdS_3 \rightarrow - S^3 \, , \quad \quad S^3 \rightarrow - AdS_3 \, .
\end{equation}
In order to get a spacetime with the correct signature the $u$, $\hat{h}_4$, $h_8$ functions need to be also analytically continued, as follows
\begin{equation}
u \rightarrow -i u \, , \quad \quad \quad \hat{h}_4 \rightarrow i \hat{h}_4 \, , \quad \quad \quad h_8 \rightarrow i h_8 \, ,\label{traca}
\end{equation}
together with $\rho \rightarrow i \rho$. 

Applying this set of transformations to the solutions to M-theory discussed in \eqref{M-theory}-\eqref{M4flux} gives rise to the following new solutions
%
%
%
\begin{eqnarray}
\text{d} s^2_{11}&=&\frac{h_8^2}{\Delta^2} \text{d} s^2_{AdS_3/\mathbb{Z}_k}+\Delta \left(\frac{u}{\sqrt{\hat{h}_4 h_8}} \text{d} s^2_{S^3}+\sqrt{\frac{\hat{h}_4}{h_8}} \text{d} s^2_{\text{CY}_2}+\frac{\sqrt{\hat{h}_4 h_8}}{u} \text{d} \rho^2\right)\label{Mmetricbis}\\
G_{(4)}&=& - \text{d}\left(-\frac{uu'}{2\hat{h}_4}+2\rho h_8\right)\wedge \hat{\text{vol}}_{\text{S}_3} - 2h_8\;\text{d} \left(\rho+\frac{u u'}{4\hat{h}_4h_8-u'^2}\right)\wedge \hat{\text{vol}}_{\text{AdS}_3/\mathds{Z}_k}\nonumber \label{M4fluxbis}\\
&&-\partial_{\rho}\hat{h}_4\; \hat{\text{vol}}_{\text{CY}_2}\\
 \Delta &=&\frac{h_8^{1/2}(4\hat{h}_4h_8-u'^2)^{1/3}}{2^{2/3}\hat{h}_4^{1/6}u^{1/3}}\label{dddelta}
\end{eqnarray}
where $k=h_8$ and the quotiented AdS$_3$ subspace is written as a Hopf fibration over AdS$_2$,
\begin{eqnarray}\label{vaza}
\text{d} s^2_{\text{AdS}_3/\mathds{Z}_k} = \frac{1}{4}\left[\left(\frac{\text{d}z}{k}+\eta\right)^2 + \text{d} s^2_{\text{AdS}_2}\right]\qquad\text{with}\qquad \text{d} \eta= \hat{\text{vol}}_{\text{AdS}_2}.\label{nara}
\end{eqnarray}

\subsection{Dual quantum mechanics}

\paragraph{} Due to the momentum charge, the previous class of solutions is dual to a 1d superconformal quantum mechanics (SCQM). Holographically, they are thus essentially different  from the AdS$_3\times $S$^3/\mathbb{Z}_k$ solutions on which the double analytic continuation was performed. From the isometries of the background, we see that the $\mathcal{N} = 4$ SCQM must preserve $\mathfrak{su}(1,1|2)$ superconformal algebra, whose bosonic sub-algebra is $\mathfrak{sl}(2) \oplus \mathfrak{su}(2)$ \cite{Frappat:1996pb}\footnote{Note that one of the SU$(2)$ isometry groups of the 3-sphere is a global symmetry.
}.

A particular solution that can be used to provide some hint on the nature of the dual quantum mechanics is the one for which $ \mathcal{I} = S^1$. This is the background that follows from setting $\hat{h}_4'=u'=0$ in (\ref{Mmetricbis})-(\ref{dddelta}). This solution is associated to a M0-M2-M5' brane intersection, and is the uplift to M-theory of the T-dual of the AdS$_3/\mathbb{Z}_M\times$ S$^3\times$ CY$_2$ geometry that describes D1-D5 branes with $M$ units of momentum along the Hopf-fibre direction of AdS$_3$. T-dualising on the Hopf-fibre gives rise to  D0-branes, D4-branes and F1-strings, as shown in Table \ref{D0-D4-F1}, which upon uplift give the M0-M2-M5' brane intersection included in Table \ref{branesetupAdS3Mq}.
 \begin{table}[ht]
	\begin{center}
		\begin{tabular}{| l | c | c | c | c| c | c| c | c| c | c | }
			\hline		    
			& 0 & 1 & 2 & 3 & 4 & 5 & 6 & 7 & 8 & 9   \\ \hline
			D0 & x &  &  &  &  &  &   &   &   &     \\ \hline
			F1 & x &  & &   &  & x  &   &   &  & \\ \hline
			D4 & x & x & x &x  & x &   &  &  &  &  \\ \hline
		\end{tabular} 
	\end{center}
	\caption{$\frac18$-BPS intersection involving D0, D4 and F1 branes.  A $\mathcal{N}=4$ supersymmetric quantum mechanics lives in the common $x^0$ direction. $(x^1,\dots, x^4)$ span the directions on the $\text{CY}_2$. $x^5$ is the direction along the $\rho$-interval. The $(x^6,\dots, x^9)$ directions enjoy an SO$(4)$ rotational symmetry, of which an SU$(2)$ is the R-symmetry of the SU$(1,1|2)$ supergroup and another SU$(2)$ a global symmetry.}   
	\label{D0-D4-F1}	
\end{table}  
When $M=1$ supersymmetry is enhanced to 2d $(4, 4)$ and the brane intersection becomes the M2-M5' brane set-up discussed in section \ref{brane-set-up}. The associated quiver is the one depicted on the right of Figure \ref{h4constant}. Switching on momentum charge allows for a quantum mechanical description of this system within M(atrix) theory, upon taking the Sen-Seiberg limit \cite{Sen:1997we,Seiberg:1997ad}. The AdS$_3/\mathbb{Z}_M\times$ S$^3\times$ CY$_2\times$ S$^1$ solution (or its AdS$_2$ reduction to Type IIA) provides for an alternative holographic description of this quantum mechanics.

More general AdS$_3/\mathbb{Z}_k\times$ S$^3\times$ CY$_2\times$ I solutions in our class should be dual to M(atrix) quantum mechanics describing M2-M5-M5' brane intersections. M5-M5' brane intersections were discussed in this context in \cite{Kachru:1998rk}, but these give rise to 4d SCFTs in their common worldvolume, and are therefore different from the intersections considered in this paper. 
The M(atrix) theory description of the 2d SCFTs living in our M-brane intersections is currently under investigation \cite{LNRS}.

Connections between AdS$_2$ solutions and M(atrix) theory have been discussed in the literature in various contexts (see for instance \cite{Dibitetto:2019nyz,Hyun:1998bi,Hyun:1998iq,Awata:1998qy,Berkooz:1999iz}). The results in \cite{Dibitetto:2019nyz} are particularly interesting regarding our system. Indeed, 
the system depicted in Table \ref{D0-D4-F1} can be thought of as the result of adding F1-branes to the 1/4-BPS D0-D4 brane system discussed in \cite{Dibitetto:2019nyz}, dual to a AdS$_2 \times $S$^3 \times$ S$^4$ geometry foliated over an interval. The D0-D4 brane system describes in M(atrix) theory longitudinal M5-branes, in terms of a $U(k)$ gauge theory with hypermultiplets in the adjoint representation and $N$ fundamentals \cite{Aharony:1997th}. This quantum mechanics is the reduction on a circle of the quiver CFT dual to the D1-D5 system (depicted on the right of Figure 
\ref{h4constant})\footnote{In the limit in which the $\text{CY}_2$ is taken to be very large, such that the group associated to the D5-branes becomes global.}. 
Alternatively, one could think of our system in terms of a 1/4-BPS D0-F1 brane system with extra D4-branes. 1/4-BPS D0-F1 branes are dual to a AdS$_2\times $S$^7$ geometry foliated over an interval 
\cite{Cvetic:2000cj,Dibitetto:2019nyz}.
Our solutions can be interpreted in these set-ups as the fully backreacted supergravity solutions that arise when F1-strings are placed in the AdS$_2\times $S$^3\times$S$^4$ geometry dual to the D0-D4 brane system, or
D4-branes are placed in the AdS$_2 \times $S$^7 \times $I solutions dual to the D0-F1 brane system, uplifted to eleven dimensions.

\section{New AdS$_2\times $S$^3\times $CY$_2$ solutions to massive Type IIA} \label{AdS2IIA}

\paragraph{} The AdS$_3/\mathbb{Z}_k\times$ S$^3\times $CY$_2\times$ I solutions to M-theory presented in (\ref{Mmetricbis})-(\ref{M4fluxbis}), can be reduced on the Hopf-fibre of AdS$_3$. This  produces a new class of AdS$_2\times$ S$^3 \times$ CY$_2\times$ I solutions to massless Type IIA supergravity. These solutions are associated to D0-F1-D4-D4' brane systems, preserve four Poincar\'e supersymmetries and have an SU(2) structure. In fact, one can check that they are just the double analytic continuation of the AdS$_3\times$ S$^2\times $ CY$_2\times$ I solutions reviewed in section \ref{AdS3S2},  when restricted to the massless case. 
Therefore, these backgrounds can be  extended straightforwardly to the massive case. In this section we present this new class of solutions. We leave their detailed study to our forthcoming publication \cite{LNRS}.

Performing the analytic continuation explained in the previous section on the class of solutions given by (\ref{eq:class I background})-(\ref{eq:class I background RR})
 we find a NS sector,
\begin{equation}
\begin{split}
\text{d}s^2 &= \frac{u}{\sqrt{\hat{h}_4 h_8}} \left( \frac{\hat{h}_4 h_8}{4 \hat{h}_4 h_8 - (u')^2} \text{d}s^2_{\text{AdS}_2} + \text{d}s^2_{\text{S}^3}\right) + \sqrt{\frac{\hat{h}_4}{h_8}} \text{d}s^2_{\text{CY}_2} + \frac{\sqrt{\hat{h}_4 h_8}}{u} \text{d} \rho^2 \, , \\
e^{- \Phi}&= \frac{h_8^{3/4}}{2 \hat{h}_4^{1/4} \sqrt{u}} \sqrt{4 \hat{h}_4 h_8 - (u')^2} \, , \quad H_{(3)} = - \frac{1}{2} \text{d} \bigg( \rho + \frac{u u'}{4 \hat{h}_4 h_8 - (u')^2} \bigg) \wedge \hat{\text{vol}}_{\text{AdS$_2$}}. 
 \label{AdS2metric}
\end{split}
\end{equation}
The RR sector reads
\begin{equation}
\begin{split}
F_{(0)} &= h_8' \, , \quad F_{(2)} = 
 - \frac{1}{2} \Big( h_8 + \frac{h_8' u' u}{4 h_8 \hat{h}_4 - (u')^2} \Big) \hat{\text{vol}}_{\text{AdS$_2$}} \, , \\
F_{(4)} &= \left( - \text{d} \bigg( \frac{u'u}{2 \hat{h}_4} \bigg) + 2 h_8 \text{d} \rho \right) \wedge \hat{\text{vol}}_{\text{S$^3$}} 
- \partial_{\rho} \hat{h}_4 \hat{\text{vol}}_{\text{CY$_2$}}.
 \label{AdS2fluxes}
\end{split}
\end{equation}
The background in equations (\ref{AdS2metric})-(\ref{AdS2fluxes}) solves the equations of motion provided that $u''=\hat{h}_4'' = h_8'' = 0$. The last two conditions come from the Bianchi identities for the RR sector.
Note that we must have $4 \hat{h}_4 h_8 - (u')^2 >0$, in order for the metric to be of the correct signature and the dilaton to be real. 

The Page fluxes are given by
\begin{eqnarray}\label{fluxes}
\hat{F}_{(0)} &= & F_{(0)} = h'_8 \, , \\
\hat{F}_{(2)} &=& F_{(2)} - F_{(0)} B_{(2)} =- \frac{1}{2} (h_8 - \rho h'_8) \hat{\text{vol}}_{\text{AdS$_2$}} \, , \\
\hat{F}_{(4)} &= &F_{(4)} = -\hat{h}'_4 \hat{\text{vol}}_{\text{CY$_2$}} - \left(2h_8-\Bigl(\frac{uu'}{2h_4}\Bigr)'\right) \hat{\text{vol}}_{\text{S$^3$}} \wedge \text{d}\rho \, ,\label{F4flux} \\
\hat{F}_{(6)} &= &\frac{1}{2}(\hat{h}_4 - \rho \hat{h}_4') \hat{\text{vol}}_{\text{AdS$_2$}} \wedge \hat{\text{vol}}_{\text{CY$_2$}} + \left( \Bigl(\frac{u(\rho u'-u)}{4h_4}\Bigr)' - \rho  h_8 \right) \hat{\text{vol}}_{\text{AdS$_2$}} \wedge \hat{\text{vol}}_{\text{S$^3$}} \wedge \mathrm{d} \rho \, , \\
\hat{F}_{(8)} &= &\left( 2 \hat{h}_4 -\Bigl(\frac{uu'}{2h_8}\Bigr)' \right) \hat{\text{vol}}_{\text{CY$_2$}} \wedge \hat{\text{vol}}_{\text{S$^3$}} \wedge \text{d} \rho \, . 
\end{eqnarray}

The class of solutions given by (\ref{AdS2metric}) and (\ref{AdS2fluxes}) provide a new class of backgrounds to Type IIA with four Poincar\'e supersymmetries and SU(2)-structure, which are warped products of AdS$_2\times $S$^3 \times$ M$_4$ over an interval, with M$_4$ a Calabi-Yau 2-fold.
The AdS$_2\times$ S$^3$ subspace realises an $\text{SL}(2, \mathbb{R}) \times \text{SO}(4)$ isometry group. As mentioned above, one of the SU$(2)$'s in $\text{SO}(4) \cong \text{SU}(2) \times \text{SU}(2)$ is a global symmetry, so the R-symmetry is that of the SU$(1,1|2)$  supergroup. 
This identifies the superconformal group of the associated dual quantum mechanics. 
As in section \ref{AdS3S2},
we have restricted ourselves to the case in which the symmetries of the $\text{CY}_2$  are respected by the full solution. We will construct the most general  class of solutions with SU(2) structure in Appendix \ref{appendix3}, where we
will relax this condition on the class I solutions in \cite{Lozano:2019emq} and analytically continue the solutions in class II.
Note that a more general class of solutions with the same supersymmetry can in principle be obtained taking an identity structure instead of the SU(2)-structure considered here.

\subsection{Brane set-up}

\paragraph{} Associated to the Page fluxes we find the following quantised charges,
\begin{equation}\label{D8 charges}
N_{8} =  2 \pi \int_{\mathcal{I}_{\rho}} \mathrm{d} \hat{F}_{(0)} \, ,  \quad \quad \mathrm{d} \hat{F}_{(0)} = h_8'' \mathrm{d} \rho \, .
\end{equation}
 According to \eqref{D8 charges}, we have a natural definition of D8-branes as objects localised in the $\rho$ direction. This, in turn, leads to the fact that D8-branes are not dissolved into fluxes, and effectively behave as flavour branes. They span the AdS$_2 \times$ S$^3 \times$ CY$_2$ sub-manifold.

From the expression for $\hat{F}_{(4)}$ in (\ref{F4flux}) we identify two manifolds supporting $\hat{F}_{(4)}$-fluxes. These are $\tilde{\mathcal{M}}_4 =\,$CY$_2$ and $\mathcal{M}_4 = S^3 \times \mathcal{I}$. There are therefore two quantised charges, associated to D4 and D4' branes
\begin{equation}\label{D4 charges}
\begin{split}
N_{4} &= \frac{\text{vol}_{\text{CY}_2}}{(2 \pi)^3} \, \int_{\mathcal{I}} \mathrm{d} \rho \,  \hat{h}_4''  \, , \\
N_{4'} &= \frac{\text{vol}_{S^3}}{(2 \pi)^3} \int_{\mathcal{I}} \mathrm{d} \rho \bigg( 2 h_8 -\Bigl(\frac{uu'}{2h_4}\Bigr)'  \bigg) \, .
\end{split}
\end{equation}
Given that
\begin{equation}
\mathrm{d} \hat{F}_{(4)} = \hat{h}_4'' \, \mathrm{d} \rho \wedge \hat{\text{vol}}_{\text{CY}_2} \, ,
\end{equation}
the D4-branes provide localised sources, and are therefore flavour branes. They are extended on  AdS$_2 \times $S$^3$. In turn, the D4'-branes are dissolved into fluxes and therefore do not provide additional physical sources. They are colour branes and extend on $(t, \text{CY}_2)$. 

Finally, there is D0 brane charge,
\begin{equation}\label{D0 charges}
N_{0} = \frac{1}{(2 \pi)^7} \text{vol}_{\text{CY}_2} \text{vol}_{S^3} \int_{\mathcal{I}} \text{d} \rho \left( 2 \hat{h}_4 -\Bigl(\frac{uu'}{2h_8}\Bigr)' \right) .
\end{equation}
Given that $\mathrm{d} \hat{F}_8$ vanishes identically the D0-branes are colour branes.
On top of this there are F1-strings, associated to the electric components of $H_{(3)}$, in (\ref{AdS2metric}). These F1 extend on AdS$_2$.

The brane set-up associated to the quantised charges is summarised in Table \ref{branesetupAdS2}. Note that this is exactly what is obtained reducing the M-brane configuration in the previous section,  with the addition of extra D8-branes, not present in M-theory.
 \begin{table}[ht]
	\begin{center}
		\begin{tabular}{| l | c | c | c | c| c | c| c | c| c | c | }
			\hline		    
			& 0 & 1 & 2 & 3 & 4 & 5 & 6 & 7 & 8 & 9  \\ \hline
			D0 & x &  &  &  &  &  &   &   &   &     \\ \hline
			F1 & x &  & &  &  & x  &   &   &   &   \\ \hline
			D4 & x &  &  &   &   &  & x & x & x & x \\ \hline
			D4' & x  &x  & x & x & x  &   &   &  &  &  \\ \hline
			D8  & x & x & x & x & x  & & x & x & x & x \\ \hline
		\end{tabular} 
	\end{center}
	\caption{$\frac18$-BPS brane intersection underlying the AdS$_2\times $S$^3$ solutions in Type IIA. $(x^1, \dots, x^4)$ span the CY$_2$ and $(x^6,x^7,x^8,x^9)$ are the transverse directions on which the SO(4) symmetry group is realised. A $\mathcal{N}=4$ supersymmetric quantum mechanics lives in the common $x^0$ direction.}  
	\label{branesetupAdS2}
\end{table} 
Similar brane intersections have been discussed in \cite{Dibitetto:2018gtk}, in connection with AdS$_2\times$ S$^3\times $S$^3$ geometries warped over a strip. In this reference the dual SCQM was interpreted in terms of D0-F1-D4 brane defects inside the 5d $Sp(N)$ fixed point theory dual to the AdS$_6$ Brandhuber-Oz background \cite{Brandhuber:1999np}. It is likely that a similar interpretation is at place for our brane system \cite{LNRS}.

\section{Conclusions}

\paragraph{} In this paper we have presented and studied new families of solutions preserving four  Poincar\'e supersymmetries.

The first is a new class of solutions to M-theory preserving ${\cal N}=(0,4)$ (small) supersymmetry, of the type AdS$_3\times$ S$^3/\mathbb{Z}_k\times$ CY$_2$ foliated over an interval. These solutions are holographic duals to 2d (small) ${\cal N}=(0,4)$  SCFTs supported by $M_A$-strings in M5-brane intersections. 
We identified the precise quivers in correspondence with the functions defining the backgrounds.
Through the computation of the central charge we have checked the compliance with the holographic dictionary and the identification of $M_A$-strings as the defining degrees of freedom of our theories.
 
 Through analytic continuation, we have constructed a second family of new solutions for which the modding is performed on the Hopf fibre of AdS$_3$. These solutions are holographically dual to SCQM, which are the M(atrix) theory descriptions of M2-M5-M5' brane intersections, upon Sen-Seiberg limit. We have postponed a more detailed analysis of these supersymmetric quantum mechanics to our forthcoming publication \cite{LNRS}. We have shown that the subclasses of AdS$_3 \times $S$^3/\mathbb{Z}_k$ and AdS$_3/\mathbb{Z}_k\times$ S$^3$ M-theory solutions with $u'=0$, contain self-dual strings. Therefore, they provide with explicit fully backreacted AdS$_3 \times$ S$^3$ OM-theory \cite{Gopakumar:2000ep} backgrounds 
\cite{Berman:2001fs}. 

Upon reduction, we have constructed a third new class of solutions to Type IIA supergravity with four Poincar\'e supercharges, of the type AdS$_2\times$ S$^3\times$ CY$_2$ foliated over an interval. These solutions should be holographic duals to the quantum mechanical systems described above, in the regime of validity of the Type IIA description. We have  extended these solutions to the massive case noticing that they are related through analytic continuation to the AdS$_3\times $S$^2\times$ CY$_2$ solutions to massive Type IIA supergravity constructed in  \cite{Lozano:2019emq}. The dual quantum mechanics is under investigation in \cite{LNRS}.
%
%
%
Three appendices extend our solutions to the most general class of AdS$_3\times $S$^2$ solutions to M-theory with (0,4) supersymmetries and SU(2) structure, and to new AdS$_2\times $S$^3\times $M$_4$ solutions to massive IIA where M$_4$ is a K\"ahler manifold. It would be interesting to understand the holographic duals for the general backgrounds presented there. 

There is an interesting connection between our work and holographic duals of defect CFTs constructed in the literature.
It was shown in \cite{Lozano:2019ywa} that a subclass of the solutions in \cite{Lozano:2019emq} allowed for an interpretation in terms of 2d D2-D4 defects in the 6d $(1,0)$ CFTs living in D6-NS5-D8 brane intersections.  
Key to this realisation was the identification of a mapping  between these solutions and the AdS$_7$ solutions to massive Type IIA supergravity constructed in \cite{Apruzzi:2013yva}. 
In the same vein, one would expect that a similar interpretation should be possible for our AdS$_3$ M-theory solutions, this time in terms of 2d M2-M5 defects in the 6d (1,0) CFTs living in M5'-branes probing $A$-type singularities. In this direction, it would be interesting to show whether our solutions bear any relation to the flows constructed in \cite{Dibitetto:2017tve,Dibitetto:2017klx}, which are asymptotically locally AdS$_7$ in the UV.   Along closely related lines, AdS$_2\times $S$^3\times$ R flows to asymptotically locally AdS$_6$ in the UV have been constructed in  \cite{Dibitetto:2018iar,Chen:2019qib}, and interpreted as 1d defects in the 5d Sp$(N)$ fixed point theory dual to the AdS$_6$ Brandhuber-Oz solution \cite{Brandhuber:1999np}. We would expect that our AdS$_2\times$ S$^3$ solutions to massive Type IIA bear a relation to these, thus allowing for an interpretation as D0-F1-D4 brane defects within the Sp$(N)$ fixed point theory living in D4'-D8 branes.
These issues are currently under investigation  \cite{FLP}.

\section*{Acknowledgements}

We would like to thank Giuseppe Dibitetto, Niall Macpherson and Nicol\`o Petri for very useful discussions. Y.L. and A.R. are partially supported by the Spanish government grant PGC2018-096894-B-100 and by the Principado de Asturias through the grant FC-GRUPIN-IDI/2018/000174. AR is supported by CONACyT-Mexico.






\appendix

\section{Review of the general AdS$_3 \times$ S$^2 \times$ M$_4$ solutions in \cite{Lozano:2019emq}}\label{appendix1}

\noindent In this appendix we summarise the generic backgrounds found in  \cite{Lozano:2019emq}. These backgrounds were divided in two classes: class I, for which the M$_4$ is a Calabi-Yau 2-fold, and class II, for which M$_4$ is a general 4d K\"ahler manifold. The particular case in class I in which
the full solution respects the symmetries of the Calabi-Yau manifold, and therefore the Calabi-Yau manifold needs to be compact, was discussed in section \ref{AdS3S2}. This is the case that concerned us in the main body of the paper. In these appendices we complete the analysis by providing the most general solutions in M-theory with (0,4) supersymmetries and SU(2) structure. In appendix \ref{appendix2} we present the uplift to M-theory of the most general solutions in class I and of the solutions in class II. In appendix \ref{appendix3} we construct AdS$_2\times $S$^3\times $M$_4$ solutions to massive Type IIA supergravity through double analytical continuation of the class I and class II solutions. 

\noindent We start reviewing the most general class I backgrounds in  \cite{Lozano:2019emq}.

\vspace{0.2cm}
\noindent\textbf{Class I: M$_4=$CY$_2$} 

\noindent The explicit form of the NS sector of the solutions referred to as class I in  \cite{Lozano:2019emq} is given by,  
\begin{equation}\label{eq:class I backgroundap}
\begin{split}
\text{d}s^2 &= \frac{u}{\sqrt{\hat{h}_4 h_8}} \left( \text{d}s^2_{\text{AdS}_3} + \frac{\hat{h}_4 h_8}{4 \hat{h}_4 h_8 + (u')^2} \text{d}s^2_{S^2} \right) + \sqrt{\frac{\hat{h}_4}{h_8}} \text{d}s^2_{\text{CY}_2} + \frac{\sqrt{\hat{h}_4 h_8}}{u} \text{d} \rho^2 \, , \\
e^{- \Phi}&= \frac{h_8^{3/4}}{2 \hat{h}_4^{1/4} \sqrt{u}} \sqrt{4 \hat{h}_4 h_8 + (u')^2} \, , \quad H_{(3)} = \frac{1}{2} \text{d} \bigg( - \rho + \frac{u u'}{4 \hat{h}_4 h_8 + (u')^2} \bigg) \wedge \widehat{\text{vol}}_{\text{S$^2$}}+ \frac{1}{h_8} \text{d} \rho \wedge H_2  \, .
\end{split}
\end{equation}
Here $\Phi$ is the dilaton, $H_{(3)}$ the NS three-form and the metric is given in string frame.  A prime denotes a derivative with respect to $\rho$. The two-form $H_2$ is defined on the CY$_2$ as we specify below.
The RR sector reads
\begin{equation}\label{eq:class I background RRap}
\begin{split}
F_{(0)} &= h_8' \, , \quad F_{(2)} =-H_2  - \frac{1}{2} \Big( h_8 - \frac{h_8' u' u}{4 h_8 \hat{h}_4+(u')^2} \Big) \hat{\text{vol}}_{\text{S$^2$}} \, , \\
F_{(4)} &= -\left( \text{d} \bigg( \frac{u'u}{2 \hat{h}_4} \bigg) + 2 h_8 \text{d} \rho \right) \wedge \hat{\text{vol}}_{\text{AdS$_3$}}   - \partial_{\rho} \hat{h}_4 \hat{\text{vol}}_{\text{CY$_2$}}- \frac{h_8}{u} (\hat{\star}_4 \text{d}_4 \hat{h}_4) \wedge d \rho \\&- \frac{u' u}{2(4 \hat{h}_4 h_8 + (u')^2)} H_2 \wedge \hat{\text{vol}}_{\text{S$^2$}}   \, ,
\end{split}
\end{equation}
where $\widehat{\star}_4$ is the Hodge dual on the CY$_2$.
Higher RR fluxes are related to these as $F_{(6)} = - \star F_{(4)}$, $F_{(8)} = \star F_{(2)}$, $F_{(10)} = - \star F_{(0)}$, where $\star$ is the ten-dimensional Hodge-dual operator.
Supersymmetry holds when
\begin{equation}\label{supersymmetry conditions}
u'' = 0 \, ,\qquad H_2+\widehat{\star}_4H_2=0\, ,
\end{equation}
which makes $u$ a linear function of $\rho$. 
$H_2$ is defined in terms of three  functions $g_{1,2,3}$ on  the CY$_2$ and the vielbein on M$_4$,  $\widehat{e}^i$,
\begin{eqnarray} \label{ges}
H_2=g_1(\hat{e}^1\wedge \hat{e}^2-\hat{e}^3\wedge \hat{e}^4)+g_2(\hat{e}^1\wedge \hat{e}^3+\hat{e}^2\wedge \hat{e}^4)+g_3(\hat{e}^1\wedge \hat{e}^4-\hat{e}^2\wedge \hat{e}^3)	.
\end{eqnarray}
Hence, the Bianchi identities of the fluxes impose
\begin{eqnarray}
\begin{split}
&h_8''=0\, , \qquad dH_2=0 \\
&\frac{h_8}{u}\nabla_{\text{CY}_2}^2\hat{h}_4+\partial_\rho^2\hat{h}_4+\frac{2}{h_8^3}(g_1^2+g_2^2+g_3^2)=0.
\end{split}
\end{eqnarray}
In the particular case when  $H_2$ vanishes and $\widehat{h}_4$  has support on the $\rho$ coordinate we find that
the supersymmetry and Bianchi identities are  satisfied for $u$, $h_8$, $\hat{h}_4$ arbitrary linear functions in $\rho$.  We are then in the case reviewed in section \ref{AdS3S2} and lifted to eleven dimensions in section \ref{Mtheoryuplift}.

\vspace{0.2cm}
\noindent\textbf{Class II: M$_4=$ K\"ahler}

\noindent We now summarise the details of the class II backgrounds in \cite{Lozano:2019emq}. These are warped products of the form AdS$_3\times$S$^2\times$M$_4\times$I, where M$_4$ is a family of K\"ahler four-manifolds with metrics that depend on the interval coordinate $\rho$, and with an integrable complex structure that is $\rho$-independent. These solutions have the following NS sector
\begin{align}\label{eq:classIINS}
\text{d}s^2&= \frac{u}{\sqrt{h w^2- v^2}}\bigg[\text{d}s_{\text{AdS}_3}^2+\frac{h w^2- v^2}{4 (h w^2- v^2)+ (u')^2} \text{d}s_{\text{S}^2}^2\bigg] +\frac{\sqrt{h w^2- v^2}}{u}\bigg[ \frac{u}{h w}\text{d}s_{\text{M}_4}^2+ \text{d}\rho^2\bigg],\nn\\[2mm]
H_{(3)}&=\frac{1}{2}\text{d}\left(-\rho+ \frac{ u u'}{4 (h w^2- v^2)+ (u')^2}\right)\wedge \text{vol}_{\text{S}^2}+ \text{d}\left(\frac{v}{w h}\hat J\right),\nn\\[2mm]
e^{-\Phi}&= \frac{w h^{\frac{1}{2}}\sqrt{4 (h w^2- v^2)+ (u')^2}}{2 \sqrt{u} (h w^2- v^2)^{\frac{1}{4}}}.
\end{align}
The functions $u,v$ and $w$ depend on $\rho$, while $h$ has support in $\rho$ and M$_4$. $\hat{J}$ is a two-form defined on the K\"ahler manifold\footnote{The interested reader  can find a detailed explanation in \cite{Lozano:2019emq}.}. The RR fluxes are given by,
\begin{align}\label{eq:classIIRR}
F_{(0)} &= v',\nn\\[2mm]
F_{(2)}&=-\frac{w^2}{u} \text{d}\rho\wedge \hat\star_4(\text{d}_4h\wedge \hat J)-\partial_{\rho}(w\hat J)+\frac{v v'}{h w}\hat J- \frac{1}{2}\left(v- \frac{v'u u'}{4(h w^2- v^2)+ (u')^2}\right)\text{vol}_{\text{S}^2},\nn\\[2mm]
F_{(4)}&=\frac{1}{2}\text{vol}_{\text{AdS}_3}\wedge\bigg(\text{d}\left(\frac{v u u'}{h w^2- v^2}\right)+ 4 v \text{d}\rho\bigg)+\frac{v}{2h}\left(\frac{ v v'}{h w^2}-\partial_{\rho}\log(v^{-1} h w^2)\right)\hat J\wedge \hat J\nn\\[2mm]
&-\frac{v w}{ u} \text{d}\rho \wedge\hat\star_4\text{d}\log h+\frac{1}{2}\bigg(\frac{u u'}{4(h w^2-v^2)+ (u')^2}F_2+ \frac{h w^2-v^2}{h w}\hat J\bigg)\wedge \text{vol}_{\text{S}^2}.
\end{align}
Here d$_4=\partial+ \overline{\partial}$, with $\partial, \overline{\partial}$ defined as the Dolbeault operators, expressed in terms of complex coordinates on $\text{M}_4$.

Supersymmetry and the Bianchi identities (away from localised sources) hold by the following conditions,
\begin{eqnarray}
\begin{split}
\label{eq:caseIIcon1}
u''&=0,~~~~\partial_{\rho}\left(\frac{ \hat{g}^{\frac{1}{2}}}{h}\right)=0,~~~~i\partial \overline{\partial}\log h = \hat{\mathfrak{R}}\, \\
\text{and}\qquad v''&=0,~~~~2i\partial\overline{\partial}h= \partial_{\rho}^2(w\hat J)\, .
\end{split}
\end{eqnarray}
The quantity $\hat{g}$ is the determinant of the metric and $\hat{\mathfrak{R}}$ the Ricci form on $\text{M}_4$. 
\section{New AdS$_3 \times$ S$^2 \times$ M$_4$ solutions in M-theory }\label{appendix2}

In this appendix we consider the uplift to eleven dimensions of the most general solutions in class I and the solutions in class II reviewed in the previous Appendix. Our backgrounds provide the most general class of AdS$_3 \times $S$^2$ solutions in M-theory with (0,4) supersymmetries and SU(2) structure. Note that a more general class of solutions with (0,4) SUSY can in principle be obtained taking an identity structure instead of the SU(2)-structure considered here.
We will focus separately on the class I and class II backgrounds.
In both cases, conditions must be imposed to allow the lift to eleven dimensions.

\vspace{0.2cm}
\noindent \textbf{Lift of the class I backgrounds} 
\\
We consider the class I geometries first.  Imposing that $F_{(0)}=0$ to allow the lift of the solutions described by equations \eqref{eq:class I backgroundap}-\eqref{eq:class I background RRap}, we find the eleven dimensional configurations,
\begin{equation}
\begin{split}
\label{m-theory-general-ClassI}
\text{d} s_{11}^2 &= \Delta \left( \frac{u}{\sqrt{{\widehat{h}}_4 h_8}} \text{d} s^2_{\text{AdS}_3} + \sqrt{\frac{{\widehat{h}}_4}{h_8}} \text{d} s^2_{\text{CY}_2} +\frac{\sqrt{{\widehat{h}_4 h_8}}}{u} \text{d} \rho^2 \right) + \frac{h_8^2}{4\Delta^2} \left( \text{d}s^2_{\text{S}^2} +(\text{D}\tilde{\psi})^2 \right)\, , \\
G_{(4)} &= - \left( \text{d} \left(\frac{u u'}{2 \widehat{h}_4} \right) + 2 h_8 \text{d} \rho \right) \wedge \widehat{\text{vol}}_{\text{AdS}_3}  - \partial_{\rho} \widehat{h}_4 \widehat{\text{vol}}_{\text{CY}_2} - \frac{u u'}{2 (\widehat{h}_4 h_8 + (u')^2)} H_2 \wedge \widehat{\text{vol}}_{\text{S}^2} \\
&{\; \; \; \; }- \frac{h_8}{u} \star_4 \text{d}_4 \widehat{h}_4 \wedge \text{d} \rho+\frac{h_8}{2}\left[ \frac{1}{2} \text{d} \left( -\rho + \frac{u u'}{4 \widehat{h}_4 h_8 + (u')^2} \right) \wedge \widehat{\text{vol}}_{\text{S}^2} + \frac{1}{h_8} \text{d} \rho \wedge H_2  \right] \wedge \text{D}\tilde{\psi} \, ,
\end{split}
\end{equation}
where we have defined the following expressions,  
\begin{eqnarray}
\begin{split}
\label{classI-def}
H_2 &= -\text{d}\cal{A}\, ,\\
\text{D}\tilde{\psi}&=\text{d}\tilde{\psi}+\tilde{\cal{A}}+\omega\qquad\text{with}\qquad \text{d}\omega=\widehat{\text{vol}}_{S^2}\,,\\
\Delta &=\frac{h_8^{1/2}(4\hat{h}_4h_8+u'^2)^{1/3}}{2^{2/3}\hat{h}_4^{1/6}u^{1/3}}\, ,
\end{split}
\end{eqnarray}
with $\tilde{\psi}= \frac{2}{h_8} \psi$ and $\tilde{\cal{A}} = \frac{2}{h_8}\cal{A}$. 
In (\ref{classI-def}) we have  assumed that $dH_2=0$ holds globally, allowing us to globally define $H_2= -d{\cal A}$.
Notice that the connection $\tilde{\cal{A}}+\omega$ makes the fibre over the S$^2$ and the CY$_2$ non trivial. The uplift to eleven dimensions preserves the ${\cal N}=(0,4)$ supersymmetry of the Type IIA solutions,
 as well as their SU(2)-structure.

Considering a sub-class of solutions --with $H_2=0$ and $\widehat{h}_4=\widehat{h}_4(\rho)$ --we obtain AdS$_3\times $S$^3/\mathbb{Z}_k\times $CY$_2$ solutions to M-theory with $(0,4)$ (small) supersymmetry, warped over an interval. These were the solutions written in equations (\ref{Mmetric})-(\ref{M4flux}).

~\\
\textbf{Lift of the class II backgrounds}
\\
To allow for a lift to M-theory, we impose that $F_{(0)}=0$. 
Considering $v'=0$ and uplifting the solution described by equations \eqref{eq:classIINS}-\eqref{eq:classIIRR} we find,
\begin{eqnarray}
ds^2_{11}&=& \Delta\bigg[\text{d}s^2_{\text{AdS}_3}+\frac{h w^2- v^2}{hwu}\left(\text{d}s_{M_4}^2+\frac{h w}{u} \text{d}\rho^2\right)\bigg]+\frac{u^2w^2h}{4(h w^2- v^2)\Delta^2}\left[\text{d}s_{\text{S}^2}^2+\frac{v^2}{w^2h}(\text{D}\tilde{\psi})^2\right]\nn\\ \nn\\
G_{(4)}&=&\frac{1}{2}\text{vol}_{\text{AdS}_3}\wedge\bigg[\text{d}\left(\frac{v u u'}{h w^2- v^2}\right)+ 4 v\,\text{d}\rho\bigg]-\frac{v}{2h}\left(\partial_{\rho}\log(v^{-1} h w^2)\right)\hat J\wedge \hat J\nn\\[2mm]
&&-\frac{v w}{ u} \text{d}\rho \wedge\hat\star_4\text{d}\log h+\frac{1}{2}\bigg(\frac{u u'}{4(h w^2-v^2)+ (u')^2}J_2+ \frac{h w^2-v^2}{h w}\hat J\bigg)\wedge\text{vol}_{\text{S}^2}\nn\\
&&+\frac{v}{2}\left[\frac{1}{2}\text{d}\left(-\rho+ \frac{ u u'}{4 (h w^2- v^2)+ (u')^2}\right)\wedge \text{vol}_{\text{S}^2}+ \text{d}\left(\frac{v}{w h}\hat J\right)\right]\wedge\text{D}\tilde{\psi},
\label{classiimtheory}\end{eqnarray}
where we have defined the following expressions,
\begin{eqnarray}
\begin{split}
\text{D}\tilde{\psi}&=\text{d}\tilde{\psi}+\tilde{\cal{J}}+\eta\qquad\text{with}\qquad \text{d}\eta=\widehat{\text{vol}}_{S^2}\, ,\\
J_2&=\text{d}\tilde{\cal{J}} =-\frac{w^2}{u} \text{d}\rho\wedge \hat{\star}_4(\text{d}_4h\wedge \hat J)-\partial_{\rho}(w\hat J)	,\\
\Delta &=\left(\frac{uw\sqrt{h}\sqrt{4(hw^2-v^2)+u'^2}}{2(hw^2-v^2)}\right)^{2/3}.
\end{split}
\end{eqnarray}
Here $\tilde{\psi}= \frac{2}{v} \psi$ and $\tilde{\cal{J}}=\frac{2}{v}\cal{J}$. As before,
the connection $\tilde{\cal{J}}+\eta$ makes the fibre over the S$^2$ and the M$_4$ non trivial. 
In order to find this uplift we have assumed that $\text{d}J_2=0$ holds globally, allowing us to globally define $J_2=\text{d}\tilde{\cal{J}}$.

The uplift to eleven dimensions preserves the ${\cal N}=(0,4)$ supersymmetry of the Type IIA solutions,
 as well as their SU(2)-structure.


\section{New AdS$_2\times $S$^3\times $M$_4$ solutions in massive Type IIA} \label{appendix3}

Applying the set of transformations discussed around equation (\ref{traca}) to the previous M-theory solutions gives rise to  AdS$_2\times $S$^3\times $M$_4$ solutions with 4 Poincar\'e supercharges and SU(2) structure. In these solutions the AdS$_2$ is non-trivially fibrered. These solutions give upon reduction to Type IIA the double analytical continuation of the class I and class II solutions reviewed in Appendix \ref{appendix1}. Thus, by acting with these rules directly on these sets of solutions we can generalise the backgrounds to the massive case. We present these backgrounds in this Appendix. The AdS$_2\times $S$^3\times $M$_4$ M-theory solutions arise upon uplift when $F_{(0)}=0$.

~\\
\textbf{Class I backgrounds:}
\\
A sub-class of these solutions was presented in section \ref{AdS2IIA}. Here we generalise this class to the case in which there is a dependence of the fluxes on the CY$_2$.
Performing the double analytical continuation  
\begin{equation}
u \rightarrow -i u \, , \quad \quad \quad \hat{h}_4 \rightarrow i \hat{h}_4 \, , \quad \quad \quad h_8 \rightarrow i h_8 \, , \quad \quad \quad \rho\rightarrow i\rho\, ,
\end{equation}
together with 
\begin{equation}
AdS_3\rightarrow -S^3\, , \quad\quad\quad S^2\rightarrow -AdS_2\, ,
\end{equation}
in the most general solutions in class I, given by equations (\ref{eq:class I backgroundap}), (\ref{eq:class I background RRap}),
we arrive at
\begin{equation}\label{NS sector analytically continued}
\begin{split}
\text{d}s^2 &= \frac{u}{\sqrt{\hat{h}_4 h_8}} \left( \frac{\hat{h}_4 h_8}{4 \hat{h}_4 h_8 - (u')^2} \text{d}s^2_{\text{AdS}_2} + \text{d}s^2_{\text{S}^3}\right) + \sqrt{\frac{\hat{h}_4}{h_8}} \text{d}s^2_{\text{CY}_2} + \frac{\sqrt{\hat{h}_4 h_8}}{u} \text{d} \rho^2 \, , \\
e^{- \Phi}&= \frac{h_8^{3/4}}{2 \hat{h}_4^{1/4} \sqrt{u}} \sqrt{4 \hat{h}_4 h_8 - (u')^2} \, , \quad H_{(3)} = - \frac{1}{2} \text{d} \bigg( \rho + \frac{u u'}{4 \hat{h}_4 h_8 - (u')^2} \bigg) \wedge \hat{\text{vol}}_{\text{AdS$_2$}} + \frac{1}{h_8} \text{d} \rho \wedge H_2 \, .
\end{split}
\end{equation}
The RR sector reads
\begin{equation}\label{RR sector analytically continued}
\begin{split}
F_{(0)} &= h_8' \, , \quad F_{(2)} = - H_2 - \frac{1}{2} \Big( h_8 + \frac{h_8' u' u}{4 h_8 \hat{h}_4 - (u')^2} \Big) \hat{\text{vol}}_{\text{AdS$_2$}} \, , \\
F_{(4)} &= \left( - \text{d} \bigg( \frac{u'u}{2 \hat{h}_4} \bigg) + 2 h_8 \text{d} \rho \right) \wedge \hat{\text{vol}}_{\text{S$^3$}}  - \frac{h_8}{u} \hat{\star}_4 \text{d}_4 h_4 \wedge d \rho - \partial_{\rho} \hat{h}_4 \hat{\text{vol}}_{\text{CY$_2$}} + \frac{u' u}{2(4 \hat{h}_4 h_8 - (u')^2)} H_2 \wedge \hat{\text{vol}}_{\text{AdS$_2$}} \, .
\end{split}
\end{equation}
These backgrounds generalise the solutions in section \ref{AdS2IIA} to the case in which $H_2\neq 0$ and $\nabla_{\text{CY}_2}\hat{h}_4\neq 0$.

Supersymmetry holds when
\begin{equation}\label{supersymmetry conditions}
u'' = 0 \, ,\qquad H_2+\widehat{\star}_4H_2=0\, .
\end{equation}
$H_2$ is defined in terms of three  functions $g_{1,2,3}$ on  the CY$_2$ and the vielbein on M$_4$,  $\widehat{e}^i$, as in eq. (\ref{ges}). 
The Bianchi identities of the fluxes impose
\begin{eqnarray}
\begin{split}
&h_8''=0\, , \qquad dH_2=0 \\
&\frac{h_8}{u}\nabla_{\text{CY}_2}^2\hat{h}_4+\partial_\rho^2\hat{h}_4+\frac{2}{h_8^3}(g_1^2+g_2^2+g_3^2)=0.
\end{split}
\end{eqnarray}
Note that it must be that $4 \hat{h}_4 h_8 - (u')^2 >0$, in order for the metric to be of the correct signature and the dilaton to be real.

~\\
\textbf{Class II backgrounds:}
\\
In this case we consider the following analytical continuation
\begin{equation}
u \rightarrow -i u \, , \quad \quad v \rightarrow i v \, , \quad w\rightarrow i w \, , \quad \quad \rho \rightarrow i \rho \, 
\end{equation}
together with 
\begin{equation}
AdS_3\rightarrow -S^3\, , \quad\quad\quad S^2\rightarrow -AdS_2\, ,
\end{equation}
of the class II solutions reviewed in Appendix \ref{appendix1}. The NS sector of the background we get reads
\begin{equation}
\begin{split}
\text{d}s^2 &= \frac{u}{\sqrt{h w^2 - v^2}} \left[ \frac{h w^2 - v^2}{4(h w^2 - v^2) - (u')^2} \text{d}s^2_{\text{AdS}_2} + \text{d}s^2_{\text{S}^3} \right] + \frac{\sqrt{h w^2 - v^2}}{u} \left[ \frac{u}{hw} \text{d}s^2_{\text{M}_4} + \text{d} \rho^2\right] \, , \\
e^{- \Phi} &= \frac{w h^{1/2} \sqrt{4(h w^2 - v^2) - (u')^2}}{2 u^{1/2} (h w^2 - v^2)^{1/4}} \, , \quad H_{(3)} = \frac{1}{2} \text{d} \left( - \rho - \frac{u u'}{4(h w^2 - v^2) - (u')^2} \right) \wedge \widehat{\text{vol}}_{\text{AdS}_2} \, . \\
\end{split}
\end{equation} 
The RR sector is given by
\begin{equation}
\begin{split}
F_{(0)} &= v' \, , \\
F_{(2)}&= - \frac{w^2}{u} \text{d} \rho \wedge \widehat{\star}_4 ( \text{d}_4 h \wedge \widehat{J}) - \partial_{\rho} w \widehat{J} + \frac{v v'}{h w} \widehat{J} - \frac{1}{2} \left( v + \frac{v' u u'}{4 (h w^2 -v^2) - (u')^2}  \right) \wedge \widehat{\text{vol}}_{\text{AdS}_2} \, , \\
F_{(4)}&= - \frac{1}{2} \widehat{\text{vol}}_{\text{S}^3} \wedge \left( \text{d} \left( \frac{v u u'}{h w^2 - v^2} \right) - 4 v \text{d} \rho \right) + \frac{v}{2 h} \left( \frac{v v'}{h w^2} - \partial_{\rho} \log (v^{-1} h w^2) \right) \widehat{J} \wedge \widehat{J} \\
&{\; \; \; \; }- \frac{v w}{u} \text{d} \rho \wedge \widehat{\star}_4 \text{d} \log h + \frac{1}{2} \left( - \frac{u u'}{4 (h w^2 - v^2) - (u')^2} F_{(2)} + \frac{h w^2 - v^2}{h w} \widehat{J} \right) \wedge \widehat{\text{vol}}_{\text{AdS}_2} \, .
\end{split}
\end{equation}
Here d$_4=\partial+ \overline{\partial}$, with $\partial, \overline{\partial}$ defined as the Dolbeault operators, expressed in terms of complex coordinates on $\text{M}_4$.
Supersymmetry and the Bianchi identities hold by the conditions,
\begin{eqnarray}
\begin{split}
\label{eq:caseIIcon1}
u''&=0,~~~~\partial_{\rho}\left(\frac{ \hat{g}^{\frac{1}{2}}}{h}\right)=0,~~~~i\partial \overline{\partial}\log h = \hat{\mathfrak{R}}\, \\
\text{and}\qquad v''&=0,~~~~2i\partial\overline{\partial}h= \partial_{\rho}^2(w\hat J)\, .
\end{split}
\end{eqnarray}

The backgrounds presented in this Appendix provide the most general class of $AdS_2\times S^3$ solutions to massive Type IIA supergravity with 4 Poincar\'e supercharges and SU(2) structure.


\begin{thebibliography}{99}


\bibitem{Maldacena:1997de}
  J.~M.~Maldacena, A.~Strominger and E.~Witten,
  ``Black hole entropy in M theory,''
  JHEP {\bf 9712} (1997) 002
  [hep-th/9711053].

\bibitem{Vafa:1997gr}
  C.~Vafa,
  ``Black holes and Calabi-Yau threefolds,''
  Adv.\ Theor.\ Math.\ Phys.\  {\bf 2} (1998) 207
  [hep-th/9711067].

\bibitem{Minasian:1999qn}
  R.~Minasian, G.~W.~Moore and D.~Tsimpis,
  ``Calabi-Yau black holes and (0,4) sigma models,''
  Commun.\ Math.\ Phys.\  {\bf 209} (2000) 325
  [hep-th/9904217].
  

\bibitem{Castro:2008ne}
  A.~Castro, J.~L.~Davis, P.~Kraus and F.~Larsen,
  ``String Theory Effects on Five-Dimensional Black Hole Physics,''
  Int.\ J.\ Mod.\ Phys.\ A {\bf 23} (2008) 613
  [arXiv:0801.1863 [hep-th]].

\bibitem{Haghighat:2015ega}
  B.~Haghighat, S.~Murthy, C.~Vafa and S.~Vandoren,
  ``F-Theory, Spinning Black Holes and Multi-string Branches,''
  JHEP {\bf 1601} (2016) 009
  [arXiv:1509.00455 [hep-th]].

\bibitem{Couzens:2019wls}
  C.~Couzens, H.~h.~Lam, K.~Mayer and S.~Vandoren,
  ``Black Holes and (0,4) SCFTs from Type IIB on K3,''
  arXiv:1904.05361 [hep-th].
 


\bibitem{Haghighat:2013tka}
B.~Haghighat, C.~Kozcaz, G.~Lockhart and C.~Vafa,
``Orbifolds of M-strings,''
Phys.\ Rev.\ D \textbf{89} (2014) no.4, 046003
[arXiv:1310.1185 [hep-th]].

\bibitem{Gadde:2015tra}
A.~Gadde, B.~Haghighat, J.~Kim, S.~Kim, G.~Lockhart and C.~Vafa,
``6d String Chains,''
JHEP \textbf{02} (2018), 143
[arXiv:1504.04614 [hep-th]].


\bibitem{Haghighat:2013gba}
B.~Haghighat, A.~Iqbal, C.~Kozçaz, G.~Lockhart and C.~Vafa,
``M-Strings,''
Commun.\ Math.\ Phys.\  \textbf{334} (2015) no.2, 779-842
[arXiv:1305.6322 [hep-th]].


 
\bibitem{Lawrie:2016axq}
  C.~Lawrie, S.~Schafer-Nameki and T.~Weigand,
  ``Chiral 2d theories from N = 4 SYM with varying coupling,''
  JHEP {\bf 1704} (2017) 111
  [arXiv:1612.05640 [hep-th]].

  	
\bibitem{Couzens:2017way}
  C.~Couzens, C.~Lawrie, D.~Martelli, S.~Schafer-Nameki and J.~M.~Wong,
  ``F-theory and AdS$_{3}$/CFT$_{2}$,''
  JHEP {\bf 1708} (2017) 043
  [arXiv:1705.04679 [hep-th]].


\bibitem{Maldacena:1997re}
  J.~M.~Maldacena,
 ``The Large N limit of superconformal field theories and supergravity,''
  Int.\ J.\ Theor.\ Phys.\  {\bf 38} (1999) 1113
   [Adv.\ Theor.\ Math.\ Phys.\  {\bf 2} (1998) 231]
  [hep-th/9711200].
	


\bibitem{Kutasov:1998zh}
  D.~Kutasov, F.~Larsen and R.~G.~Leigh,
  ``String theory in magnetic monopole backgrounds,''
  Nucl.\ Phys.\ B {\bf 550} (1999) 183
  [hep-th/9812027].

\bibitem{Sugawara:1999qp}
  Y.~Sugawara,
  ``N = (0,4) quiver SCFT(2) and supergravity on AdS(3) x S**2,''
  JHEP {\bf 9906} (1999) 035
  [hep-th/9903120].

\bibitem{Larsen:1999dh}
  F.~Larsen and E.~J.~Martinec,
  ``Currents and moduli in the (4,0) theory,''
  JHEP {\bf 9911} (1999) 002
  [hep-th/9909088].

\bibitem{Okuyama:2005gq}
  K.~Okuyama,
  ``D1-D5 on ALE space,''
  JHEP {\bf 0512} (2005) 042
  [hep-th/0510195].

\bibitem{Douglas:1996uz}
  M.~R.~Douglas,
  ``Gauge fields and D-branes,''
  J.\ Geom.\ Phys.\  {\bf 28} (1998) 255
  [hep-th/9604198].


\bibitem{Lozano:2019emq}
  Y.~Lozano, N.~T.~Macpherson, C.~Nunez and A.~Ramirez,
  ``AdS$_3$ solutions in Massive IIA with small $\mathcal{N}=(4,0)$ supersymmetry,''
  JHEP {\bf 2001} (2020) 129
  [arXiv:1908.09851 [hep-th]].

\bibitem{Kim:2005ez}
  N.~Kim,
  ``AdS(3) solutions of IIB supergravity from D3-branes,''
  JHEP {\bf 0601} (2006) 094
  [hep-th/0511029].

\bibitem{Gauntlett:2006af}
  J.~P.~Gauntlett, O.~A.~P.~Mac Conamhna, T.~Mateos and D.~Waldram,
  ``Supersymmetric AdS(3) solutions of type IIB supergravity,''
  Phys.\ Rev.\ Lett.\  {\bf 97} (2006) 171601
  [hep-th/0606221].

\bibitem{Gauntlett:2006ns}
  J.~P.~Gauntlett, N.~Kim and D.~Waldram,
  ``Supersymmetric AdS(3), AdS(2) and Bubble Solutions,''
  JHEP {\bf 0704} (2007) 005
  [hep-th/0612253].



\bibitem{DHoker:2008lup}
E.~D'Hoker, J.~Estes, M.~Gutperle and D.~Krym,
``Exact Half-BPS Flux Solutions in M-theory. I: Local Solutions,''
JHEP \textbf{08} (2008), 028
[arXiv:0806.0605 [hep-th]].

\bibitem{Donos:2008hd}
A.~Donos, J.~P.~Gauntlett and J.~Sparks,
``AdS(3) x (S**3 x S**3 x S**1) Solutions of Type IIB String Theory,''
Class. Quant. Grav. \textbf{26} (2009), 065009
[arXiv:0810.1379 [hep-th]].

\bibitem{DHoker:2008rje}
E.~D'Hoker, J.~Estes, M.~Gutperle and D.~Krym,
``Exact Half-BPS Flux Solutions in M-theory II: Global solutions asymptotic to AdS(7) x S**4,''
JHEP \textbf{12} (2008), 044
[arXiv:0810.4647 [hep-th]].

\bibitem{DHoker:2009wlx}
E.~D'Hoker, J.~Estes, M.~Gutperle and D.~Krym,
``Exact Half-BPS Flux Solutions in M-theory III: Existence and rigidity of global solutions asymptotic to AdS(4) x S**7,''
JHEP \textbf{09} (2009), 067
[arXiv:0906.0596 [hep-th]].

\bibitem{Kim:2012ek}
  N.~Kim,
  ``The Backreacted K\'ahler Geometry of Wrapped Branes,''
  Phys.\ Rev.\ D {\bf 86} (2012) 067901
  [arXiv:1206.1536 [hep-th]].
	
\bibitem{Kelekci:2016uqv}
  O.~Kelekci, Y.~Lozano, J.~Montero, E.~A.~Colgain and M.~Park,
  ``Large superconformal near-horizons from M-theory,''
  Phys.\ Rev.\ D {\bf 93} (2016) no.8,  086010
  [arXiv:1602.02802 [hep-th]].

\bibitem{Eberhardt:2017uup}
  L.~Eberhardt,
  ``Supersymmetric AdS$_{3}$ supergravity backgrounds and holography,''
  JHEP {\bf 1802} (2018) 087
  [arXiv:1710.09826 [hep-th]].

\bibitem{Couzens:2017nnr}
  C.~Couzens, D.~Martelli and S.~Schafer-Nameki,
  ``F-theory and AdS$_{3}$/CFT$_{2}$ (2, 0),''
  JHEP {\bf 1806} (2018) 008
  [arXiv:1712.07631 [hep-th]].

\bibitem{Dibitetto:2018ftj}
  G.~Dibitetto, G.~Lo Monaco, A.~Passias, N.~Petri and A.~Tomasiello,
  ``AdS$_3$ Solutions with Exceptional Supersymmetry,''
  Fortsch.\ Phys.\  {\bf 66} (2018) no.10,  1800060
  [arXiv:1807.06602 [hep-th]].
	
\bibitem{Macpherson:2018mif}
  N.~T.~Macpherson,
  ``Type II solutions on AdS$_{3} \times$ S$^{3} \times$ S$^{3}$ with large superconformal symmetry,''
  JHEP {\bf 1905} (2019) 089
  [arXiv:1812.10172 [hep-th]].
	
\bibitem{Deger:2019tem}
  N.~S.~Deger, C.~Eloy and H.~Samtleben,
  ``${\mathcal{N}=(8,0)}$ AdS vacua of three-dimensional supergravity,''
  arXiv:1907.12764 [hep-th].
	

\bibitem{Legramandi:2019xqd}
A.~Legramandi and N.~T.~Macpherson,
``AdS$_3$ solutions with $\mathcal{N}=(3,0)$ from S$^3\times$S$^3$ fibrations,''
[arXiv:1912.10509 [hep-th]].


\bibitem{Lozano:2019jza}
  Y.~Lozano, N.~T.~Macpherson, C.~Nunez and A.~Ramirez,
 ``1/4 BPS solutions and the AdS$_3$/CFT$_2$ correspondence,''
  Phys.\ Rev.\ D {\bf 101} (2020) no.2,  026014
  [arXiv:1909.09636 [hep-th]].
  
\bibitem{Lozano:2019zvg}
  Y.~Lozano, N.~T.~Macpherson, C.~Nunez and A.~Ramirez,
  ``Two dimensional ${\cal N}=(0,4)$ quivers dual to AdS$_3$ solutions in massive IIA,''
  JHEP {\bf 2001} (2020) 140
  [arXiv:1909.10510 [hep-th]].

	
\bibitem{Colgain:2010wb}
  E.~O Colgain, J.~B.~Wu and H.~Yavartanoo,
  ``Supersymmetric AdS3 X S2 M-theory geometries with fluxes,''
  JHEP {\bf 1008} (2010) 114
  [arXiv:1005.4527 [hep-th]].


\bibitem{Berkooz:1996is} 
  M.~Berkooz and M.~R.~Douglas,
  ``Five-branes in M(atrix) theory,''
  Phys.\ Lett.\ B {\bf 395}, 196 (1997)
  [hep-th/9610236].


\bibitem{Aharony:1996bh} 
  O.~Aharony and M.~Berkooz,
  ``Membrane dynamics in M(atrix) theory,''
  Nucl.\ Phys.\ B {\bf 491}, 184 (1997)
  [hep-th/9611215].

\bibitem{Aharony:1997th} 
  O.~Aharony, M.~Berkooz, S.~Kachru, N.~Seiberg and E.~Silverstein,
  ``Matrix description of interacting theories in six-dimensions,''
  Adv.\ Theor.\ Math.\ Phys.\  {\bf 1}, 148 (1998)
  [hep-th/9707079].


\bibitem{Aharony:1997an}
  O.~Aharony, M.~Berkooz and N.~Seiberg,
  ``Light cone description of (2,0) superconformal theories in six-dimensions,''
  Adv.\ Theor.\ Math.\ Phys.\  {\bf 2} (1998) 119
  [hep-th/9712117].

\bibitem{Hanany:1997xc} 
  A.~Hanany and G.~Lifschytz,
  ``M(atrix) theory on T**6 and a m(atrix) theory description of K K monopoles,''
  Nucl.\ Phys.\ B {\bf 519}, 195 (1998)
  [hep-th/9708037].

\bibitem{Kachru:1998rk}
S.~Kachru, Y.~Oz and Z.~Yin,
``Matrix description of intersecting M-5 branes,''
JHEP \textbf{11} (1998), 004
[arXiv:hep-th/9803050 [hep-th]].

\bibitem{FLP}
F.~Faedo, Y.~Lozano, N.~Petri, in preparation.

\bibitem{Berman:2001fs}
  D.~S.~Berman and P.~Sundell,
  ``AdS(3) OM theory and the selfdual string or membranes ending on the five -brane,''
  Phys.\ Lett.\ B {\bf 529} (2002) 171
  [hep-th/0105288].

\bibitem{Gopakumar:2000ep}
  R.~Gopakumar, S.~Minwalla, N.~Seiberg and A.~Strominger,
  ``(OM) theory in diverse dimensions,''
  JHEP {\bf 0008} (2000) 008
  [hep-th/0006062].

\bibitem{Kim:2015gha}
J.~Kim, S.~Kim and K.~Lee,
``Little strings and T-duality,''
JHEP \textbf{02} (2016), 170
[arXiv:1503.07277 [hep-th]].


\bibitem{Boonstra:1998yu}
H.~J.~Boonstra, B.~Peeters and K.~Skenderis,
``Brane intersections, anti-de Sitter space-times and dual superconformal theories,''
Nucl.\ Phys.\ B \textbf{533} (1998), 127-162
[arXiv:hep-th/9803231 [hep-th]].



\bibitem{ArkaniHamed:2001ca}
N.~Arkani-Hamed, A.~G.~Cohen and H.~Georgi,
``(De)constructing dimensions,''
Phys. Rev. Lett. \textbf{86} (2001), 4757-4761
[arXiv:hep-th/0104005 [hep-th]].

\bibitem{Constable:2002vt}
N.~R.~Constable, J.~Erdmenger, Z.~Guralnik and I.~Kirsch,
``(De)constructing intersecting M5-branes,''
Phys. Rev. D \textbf{67} (2003), 106005
[arXiv:hep-th/0212136 [hep-th]].


\bibitem{Dibitetto:2019nyz}
  G.~Dibitetto, Y.~Lozano, N.~Petri and A.~Ramirez,
  ``Holographic Description of M-branes via AdS$_2$,''
  arXiv:1912.09932 [hep-th].


\bibitem{Putrov:2015jpa}
P.~Putrov, J.~Song and W.~Yan,
``(0,4) dualities,''
JHEP \textbf{03} (2016), 185
[arXiv:1505.07110 [hep-th]].

\bibitem{Frappat:1996pb}
L.~Frappat, P.~Sorba and A.~Sciarrino,
``Dictionary on Lie superalgebras,''
[arXiv:hep-th/9607161 [hep-th]].



\bibitem{Sen:1997we}
  A.~Sen,
  ``D0-branes on T**n and matrix theory,''
  Adv.\ Theor.\ Math.\ Phys.\  {\bf 2} (1998) 51
  [hep-th/9709220].
 

\bibitem{Seiberg:1997ad}
  N.~Seiberg,
  ``Why is the matrix model correct?,''
  Phys.\ Rev.\ Lett.\  {\bf 79} (1997) 3577
  [hep-th/9710009].
  
\bibitem{LNRS}
Y.~Lozano, C.~Nunez, A.~Ramirez, S.~Speziali, in preparation.

  
\bibitem{Hyun:1998bi}
  S.~Hyun,
  ``The Background geometry of DLCQ supergravity,''
  Phys.\ Lett.\ B {\bf 441} (1998) 116
  [hep-th/9802026].

\bibitem{Hyun:1998iq}
  S.~Hyun and Y.~Kiem,
  ``Background geometry of DLCQ M theory on a p - torus and holography,''
  Phys.\ Rev.\ D {\bf 59} (1999) 026003
  [hep-th/9805136].


\bibitem{Awata:1998qy}
  H.~Awata and S.~Hirano,
  ``AdS(7) / CFT(6) correspondence and matrix models of M5-branes,''
  Adv.\ Theor.\ Math.\ Phys.\  {\bf 3} (1999) 147
  [hep-th/9812218].
  
\bibitem{Berkooz:1999iz}
  M.~Berkooz and H.~L.~Verlinde,
  ``Matrix theory, AdS / CFT and Higgs-Coulomb equivalence,''
  JHEP {\bf 9911} (1999) 037
  [hep-th/9907100].
  
\bibitem{Cvetic:2000cj} 
  M.~Cvetic, H.~Lu, C.~N.~Pope and J.~F.~Vazquez-Poritz,
  ``AdS in warped space-times,''
  Phys.\ Rev.\ D {\bf 62}, 122003 (2000)
  [hep-th/0005246].
  
\bibitem{Dibitetto:2018gtk}
G.~Dibitetto and N.~Petri,
``AdS$_{2}$ solutions and their massive IIA origin,''
JHEP \textbf{05} (2019), 107
[arXiv:1811.11572 [hep-th]].

\bibitem{Brandhuber:1999np}
A.~Brandhuber and Y.~Oz,
``The D-4 - D-8 brane system and five-dimensional fixed points,''
Phys. Lett. B \textbf{460} (1999), 307-312
[arXiv:hep-th/9905148 [hep-th]].

\bibitem{Lozano:2019ywa}
Y.~Lozano, N.~T.~Macpherson, C.~Nunez and A.~Ramirez,
``AdS$_3$ solutions in massive IIA, defect CFTs and T-duality,''
JHEP \textbf{12} (2019), 013
[arXiv:1909.11669 [hep-th]].

\bibitem{Apruzzi:2013yva}
F.~Apruzzi, M.~Fazzi, D.~Rosa and A.~Tomasiello,
``All AdS$_7$ solutions of type II supergravity,''
JHEP \textbf{04} (2014), 064
[arXiv:1309.2949 [hep-th]].
   

\bibitem{Dibitetto:2017tve} 
  G.~Dibitetto and N.~Petri,
  ``BPS objects in D = 7 supergravity and their M-theory origin,''
  JHEP {\bf 1712}, 041 (2017)
  [arXiv:1707.06152 [hep-th]].



\bibitem{Dibitetto:2017klx} 
  G.~Dibitetto and N.~Petri,
  ``6d surface defects from massive type IIA,''
  JHEP {\bf 1801}, 039 (2018)
  [arXiv:1707.06154 [hep-th]].

\bibitem{Dibitetto:2018iar} 
  G.~Dibitetto and N.~Petri,
  ``Surface defects in the D4 $-$ D8 brane system,''
  JHEP {\bf 1901}, 193 (2019)
  [arXiv:1807.07768 [hep-th]].

\bibitem{Chen:2019qib}
K.~Chen and M.~Gutperle,
``Holographic line defects in F(4) gauged supergravity,''
Phys. Rev. D \textbf{100} (2019) no.12, 126015
[arXiv:1909.11127 [hep-th]].



\end{thebibliography}
\end{document}